\PassOptionsToPackage{table}{xcolor}
\documentclass[a4paper,fleqn]{cas-dc}

\usepackage[authoryear]{natbib}

\usepackage{bm}

\def\eqref#1{equation~\ref{#1}}

\def\1{\bm{1}}

\DeclareMathAlphabet{\mathsfit}{\encodingdefault}{\sfdefault}{m}{sl}
\SetMathAlphabet{\mathsfit}{bold}{\encodingdefault}{\sfdefault}{bx}{n}

\usepackage{mathtools}
\usepackage{amsthm}
\usepackage{subcaption}
\usepackage{enumitem}
\usepackage[capitalize,noabbrev]{cleveref}
\usepackage[textsize=tiny]{todonotes}
\usepackage{xurl}

\theoremstyle{plain}
\newtheorem{theorem}{Theorem}[section]

\theoremstyle{definition}

\theoremstyle{remark}
\newtheorem{remark}[theorem]{Remark}

\begin{document}
\let\WriteBookmarks\relax
\def\floatpagepagefraction{1}
\def\textpagefraction{.001}

\shorttitle{Landslide reconstruction with LAPIS-SHRED}
\shortauthors{X. Zhang et~al.}

\title[mode=title]{Spatiotemporal Landslide Runout Reconstruction from Sparse Observations with LAPIS-SHRED}

\author[1]{Xingyue Zhang}
\cormark[1]
\ead{xyzhamy@gmail.com}
\credit{Writing -- original draft, Writing -- review \& editing, Software, Formal analysis, Visualization, Validation}

\author[2]{Yuxuan Bao}
\ead{baoyx@uw.edu}
\credit{Writing -- review \& editing, Methodology, Software, Formal analysis, Conceptualization}

\author[3]{Jeffrey Keck}
\ead{keckje@gmail.com}
\credit{Writing -- review \& editing, Data curation, Resources}

\author[4]{Erkan Istanbulluoglu}
\ead{erkani@uw.edu}
\credit{Writing -- review \& editing, Data curation, Resources, Supervision, Conceptualization, Funding acquisition}

\author[5]{J. Nathan Kutz}
\ead{kutz@uw.edu}
\credit{Writing -- review \& editing, Supervision, Conceptualization, Funding acquisition}

\affiliation[1]{organization={School of Environmental and Forest Sciences, University of Washington},
            city={Seattle},
            postcode={98195},
            state={WA},
            country={USA}}

\affiliation[2]{organization={Department of Applied Mathematics, University of Washington},
            city={Seattle},
            postcode={98195},
            state={WA},
            country={USA}}

\affiliation[3]{organization={Washington State Department of Natural Resources, Forest Resources Division},
            city={Olympia},
            postcode={98504},
            state={WA},
            country={USA}}

\affiliation[4]{organization={Department of Civil \& Environmental Engineering, University of Washington},
            city={Seattle},
            postcode={98195},
            state={WA},
            country={USA}}

\affiliation[5]{organization={Autodesk Research},
            city={London},
            country={UK}}

\cortext[cor1]{Corresponding author}

\begin{abstract}
Reconstructing the spatiotemporal evolution of landslide runout from limited observations is a central challenge in geohazard assessment: transient flow dynamics are rarely observed directly, and available data are often confined to post-event deposits or brief monitoring windows. We apply LAPIS-SHRED (LAtent Phase Inference from Short time sequence using SHallow REcurrent Decoders) to reconstruct runout events near Mount St.\ Helens, USA. Trained on physics-based simulation ensembles, the framework supports both backward reconstruction from single frame and forward prediction from short initial window across different runout regimes. Using 32 sensor observations from a single timepoint, LAPIS-SHRED recovers complete runout trajectories with NRMSE below 5.8\%, robust across varying observational and training conditions. Validation against observed deposits yields spatial correlation of $r = 0.843$ against the calibrated simulator's $r = 0.853$. These results suggest that latent-space temporal inference offers a viable pathway for spatiotemporal landslide reconstruction under extreme observational constraints typical of field settings.
\end{abstract}

\begin{keywords}
Landslide runout \sep Spatiotemporal reconstruction \sep Sparse observations \sep Data-driven machine learning
\end{keywords}

\maketitle

\section{Introduction}
\label{sec:intro}

Landslides are among the most widespread and destructive geohazards, posing significant risks to infrastructure, ecosystems, and human life in mountainous and tectonically active regions~\citep{petley2012global,froude2018global}. Understanding and predicting landslide runout behavior---including the spatial extent, flow pathways, and temporal evolution of mobilized material---is critical for hazard assessment, risk mitigation, and post-event analysis~\citep{corominas2014recommendations,hungr2014varnes,fan2019earthquake}. However, direct observation of landslide dynamics is typically limited: measurements are often sparse in space and time, and in many cases only heterogeneous post-event observations of erosion and deposition are available---captured through field surveys, aerial photography, or LiDAR differencing~\citep{guzzetti2012landslide,jaboyedoff2012use}---while the transient flow processes that govern runout remain unobserved~\citep{jakob2005debris,iverson1997physics}. In addition, post-event observations may not capture the full event mass balance when debris enters rivers, where material can be transported, submerged, or reworked~\citep{hovius2011prolonged}. Their interpretation can also be difficult for process reconstruction when multiple runouts from a cluster of landslides coalesce in valley networks. 

Physics-based runout models provide a principled framework for simulating landslide processes, ranging from depth-averaged continuum formulations~\citep{mcdougall2004model,iverson2014depth,delannay2017granular} to cellular automaton approaches that capture bulk transport behavior~\citep{keck2024landslide}. Runout models are typically calibrated after major landslide events with downstream impacts, to predict future runout at the same site or at similar sites. However, calibrated models may not perform well under future storm events due to uncertainties in watershed sediment availability, material properties, disturbance histories (e.g., wildfires, timber harvest, land development), hydrologic drivers, initiation mechanisms, and failure geometry. This leads to event-specific parameterizations, with each calibration tied to a fixed set of model parameters, thereby limiting the broader use of these models for both pre-event hazard assessment and post-event reconstruction ~\citep{mcdougall20172014,fan2019earthquake,kean2019inundation}. 

A potential alternative to forward modeling with calibrated models is inverse reconstruction, which works backward from limited observations to infer the most plausible evolution of a particular runout event, similar to the post-event back-analysis commonly used in runout modeling literature~\citep{mcdougall2004model}. Rather than using a past event calibration to predict what might occur under assumed future conditions, inverse reconstruction seeks the sequence of landslide initiation and transient flow dynamics that is most consistent with the available observations and process constraints. However, traditional data assimilation and inverse modeling approaches~\citep{tarantola2005inverse,evensen2009data} generally require either densely sampled temporal observations or computationally intensive iterative optimization, both of which are often infeasible in data-limited field settings. 

Ensemble-based inverse approaches can in principle reconstruct unobserved dynamics, but full-physics ensemble inversions are computationally expensive and scale poorly with the dimensionality of the spatiotemporal state~\citep{evensen2009data, houtekamer1998data}, while traditional surrogate models, such as polynomial chaos expansions~\citep{xiu2002wiener, sudret2008global} or Gaussian process emulators~\citep{sacks1989design, rasmussen2003gaussian}, reduce the per-evaluation cost but do not directly address the reconstruction of unobserved spatiotemporal fields under sparse sensing constraints typical of landslide monitoring.

These constraints motivate latent-space approaches that learn compressed representations of system dynamics from simulation ensembles~\citep{berkooz1993proper,benner2015survey} and can operate under severe observational sparsity. 

Recent advances in data-driven modeling have introduced new opportunities for reconstructing spatiotemporal dynamics from sparse observations~\citep{raissi2019physics,karniadakis2021physics,brunton2024promising}. In particular, latent-space methods based on neural architectures have demonstrated the ability to learn low-dimensional representations of high-dimensional physical systems (e.g., reduced representations of the dominant system behavior) and to perform reconstruction or forecasting under partial observability~\citep{erichson2020shallow,vlachas2018data,chen2018neural}. The SHallow REcurrent Decoder (SHRED) framework~\citep{williams2024sensing} leverages temporal embeddings of sparse sensor measurements to reconstruct full spatial states, while extensions such as SINDy-SHRED~\citep{gao2026sparse} and DA-SHRED~\citep{bao2025data} address equation discovery and data assimilation, respectively. 
Whereas standard SHRED and its extensions assume access to sensor histories spanning the full temporal extent of the trajectory, LAPIS-SHRED~\citep{bao2026latent} alleviates this assumption, enabling reconstruction from observation windows confined to a small fraction of the temporal data---as few as a single terminal frame.

To address the research question posed above, we investigate the application of LAPIS-SHRED to landslide runout reconstruction under extreme observational constraints, by asking whether sparse terminal or early-stage observations can support full spatiotemporal reconstruction of the unobserved evolution of landslide runout events, and how LAPIS-SHRED compares with alternative SHRED formulations under different observation regimes.
Specifically, we consider two complementary inference settings: (i)~\emph{backward inference}, in which only a terminal observation---such as a map of runout erosion and deposition---is available and the goal is to reconstruct the antecedent evolution; and (ii)~\emph{forward inference}, in which a short initial observation window is provided, such as an early-time flow stage from imagery or sensors, and the goal is to predict the subsequent runout. Both settings reflect practical limitations in landslide monitoring and runout prediction, where continuous observations of runout evolution are rarely available, and motivate a framework capable of leveraging simulation priors to infer unobserved dynamics and support rapid predictions over large areas.

We evaluate the approach outlined above on two documented landslide events within the same valley in the Cascade Mountains near Mount St.\ Helens, using ensembles generated by the Mass Wasting Runout (MWR) model~\citep{keck2024landslide}, a physics-based cellular automaton implemented within the Landlab framework~\citep{hobley2017creative,barnhart2020landlab}. The experiments examine reconstruction accuracy across varying environmental conditions, sensor configurations, digital elevation model (DEM) resolutions, and parameter perturbation regimes. The principal contributions of this work are as follows:

\begin{enumerate}[nosep]
    \item We demonstrate that latent-space temporal inference can reconstruct landslide runout dynamics from highly limited observations, including single terminal snapshots.
    \item We provide a systematic evaluation of factors affecting reconstruction performance, including terrain resolution, sensor placement, and parameter uncertainty. We also validate against actual field-observed deposits from a runout event.
    \item We discuss the implications of these findings for geohazard assessment and highlight both the potential and limitations of simulation-informed machine learning approaches for landslide reconstruction.
\end{enumerate}

The remainder of this paper is organized as follows. Section~\ref{sec:data} introduces the study site, the MWR runout model and its calibration, and the simulation ensemble generation procedure. Section~\ref{sec:method} formulates the LAPIS-SHRED framework for landslide reconstruction, including the problem formulation for backward and forward inference. Section~\ref{sec:results} presents the core reconstruction results across runout regimes, baseline comparisons, sensitivity analysis, and a preliminary validation against field-observed deposits from the 2022 event. Sections~\ref{sec:discussion} and~\ref{sec:future} discuss implications, limitations, and future directions for extending the framework to broader and more complex applications.

\section{Preliminaries}
\label{sec:data}

\begin{figure*}[t]
\centering
\includegraphics[width=\textwidth]{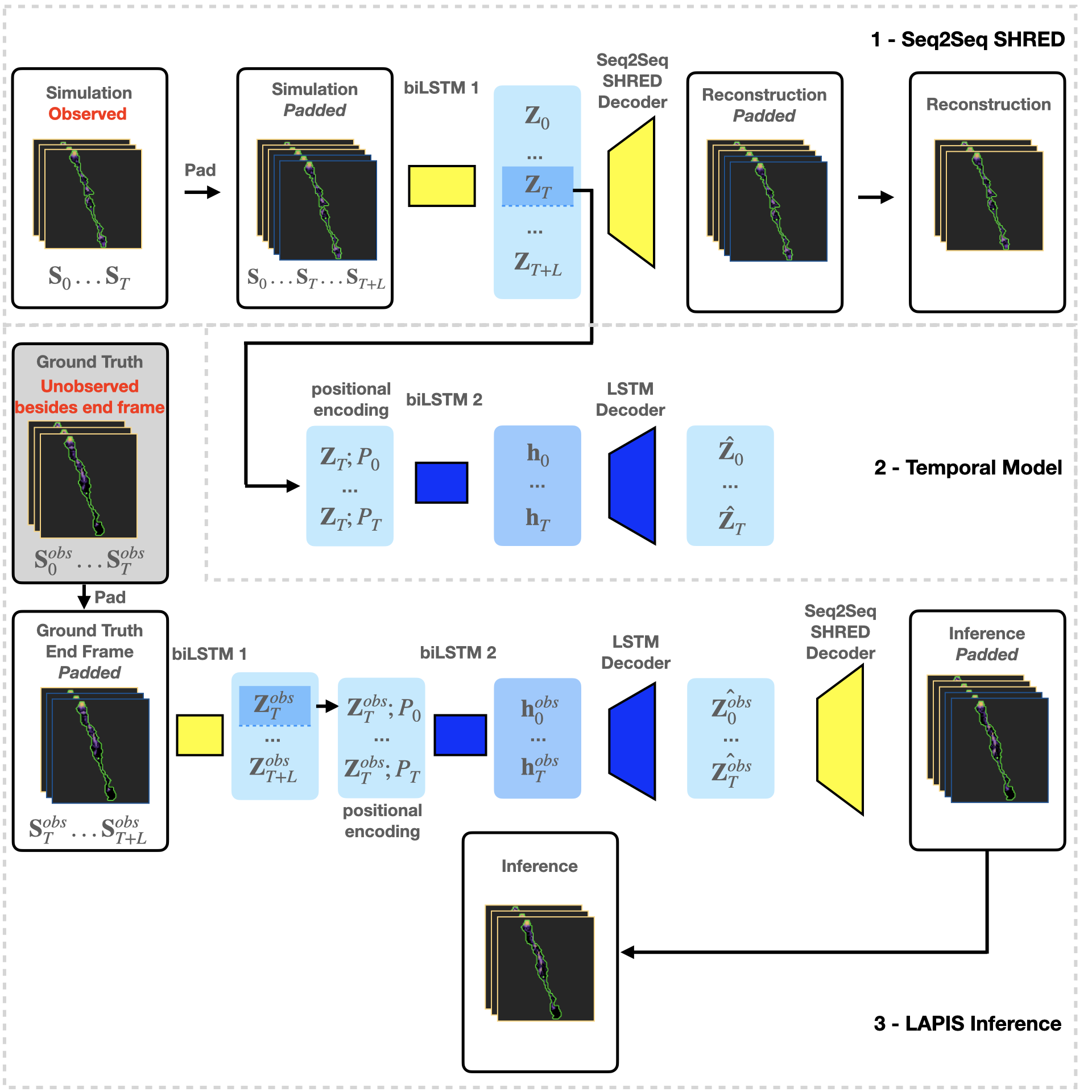}
\caption{Overview of the LAPIS architecture. \textbf{Stage 1 (Seq2Seq SHRED):} Sparse sensor observations $\mathbf{s}_0, \ldots, \mathbf{s}_T$ from simulation ensembles are padded with $L$ copies of the terminal frame and encoded via a bidirectional LSTM into latent representations $\mathbf{z}_0, \ldots, \mathbf{z}_{T+L}$, which are decoded to reconstruct the full spatiotemporal trajectory. \textbf{Stage 2 (Temporal Model):} A separate network learns to map the terminal latent $\mathbf{z}_T$ (augmented with positional encodings) to the complete preceding latent sequence $\hat{\mathbf{z}}_0, \ldots, \hat{\mathbf{z}}_T$. \textbf{Stage 3 (LAPIS Inference):} At test time, only the terminal observation $\mathbf{s}_T^{\text{obs}}$ is available. Static padding constructs a pseudo-sequence, the frozen SHRED temporal unit extracts the terminal latent $\mathbf{z}_T^{\text{obs}}$, the backward model recovers the full latent trajectory, and the frozen SHRED decoder reconstructs the complete spatiotemporal evolution.}
\label{fig:lapis_architecture}
\end{figure*}

We develop and evaluate LAPIS-SHRED on landslide runout simulation, a geophysical process characterized by rapid mass movement followed by terminal deposition. This section introduces the physics-based simulation framework and the data generation pipeline employed for training and evaluation.

\subsection{Landslide Runout Dynamics and the Mass Wasting Runout Model}
\label{sec:mwr_model}

Landslide runout processes---including debris flows, debris avalanches, and related mass wasting phenomena---involve the gravitationally-driven transport of failed hillslope material across complex terrain~\citep{iverson1997physics}. The source of sediment could be a shallow or deep-seated landslide, acceleration of initially slow-moving landslide, or slurry-type flows often driven by concentrated runoff erosion. These processes pose significant hazards to infrastructure and human life, while playing a fundamental role in the evolution of earth surface and biogeochemical cycles, motivating the development of predictive models for runout extent, erosion, and deposition~\citep{fan2019earthquake, kean2019inundation}.

Numerous physics-based models for simulating landslide runout exist, varying from detailed, continuum-based mechanistic models (e.g., D-Claw; \citealp{iverson2001flow}) to reduced complexity models (e.g., \citealp{han2021hydrodynamic}). Here, we employ the Mass Wasting Runout (MWR) component~\citep{keck2024landslide} implemented within the Landlab earth surface modeling framework~\citep{hobley2017creative, barnhart2020landlab}. MWR is a reduced-complexity, physically-informed cellular automaton model that simulates the downslope progression of mass-wasting events by enforcing mass continuity at each grid cell. At a given Landlab node $n$, incoming sediment flux $q_I$, local erosion $E$ and aggradation $A$ determine the outgoing flux $q_O$, which in turn governs runout extent and topographic change. Material is routed using a multiple-flow-direction scheme that partitions flux according to local slope. Runout transport and erosion are governed by optional approximations for quasi-static basal shear stress and grain-size-based flow shear stress, while aggradation is represented by a “spreading” rule using a critical-slope $S_c$ and flow thickness $q_c$ criteria. Relative to more complex models such as D-Claw, MWR is computationally efficient while still capable of representing a range of runout behaviors, from relatively dry, grain-dominated flows to more slurry-like, incisive flows, making it well-suited for this exploratory reconstruction study. The model output is represented here by the cumulative topographic change, $\Delta z = z(t) - z_0$, which we refer to as the Difference-of-DEMs (DoD) field and denote by $Y(t) \in \mathbb{R}^{H \times W}$ over a spatial domain of dimensions $H \times W$.

\textbf{Key Parameters.} The MWR model is governed by three calibratable parameters that control distinct aspects of the runout process:

\begin{enumerate}[nosep]
    \item \textit{Critical slope} $S_c \in (0, 1)$: The threshold terrain gradient below which deposition begins to occur and above which erosion dominates. This parameter is conceptually similar to the angle of repose of dry hillslope materials and can often be approximated from the surface slope of observed deposits.
    
    \item \textit{Threshold flux} $q_c > 0$ (m): The flux magnitude below which flow resistance (whether from basal friction, debris obstacles, or vegetation) becomes sufficient to arrest forward momentum, causing deposition. This parameter effectively controls the minimum flow depth required for continued propagation.
    
    \item \textit{Erosion coefficient} $k > 0$: Soil erodibility parameter that relates the rate of substrate entrainment to flow shear stress. Higher values produce deeper erosion channels and greater sediment bulking.
\end{enumerate}

In addition to the transport parameters, two initial-condition parameters control the soil available for entrainment:

\begin{enumerate}[nosep, resume]
    \item \textit{Soil depth} $d_{\text{soil}} > 0$ (m): The baseline soil thickness, representing the ambient regolith mantle available for erosion along runout path. This parameter is assumed constant in this study but it can be obtained from field surveys or soil thickness maps from gridded soil data sets.
    
    \item \textit{Depth parameter} $d_p > 0$ (m): Controls the deep-seated soil thickness within the failure zone at the source of the landslide.
\end{enumerate}

We denote the parameter vector as
$\theta = (S_c,\allowbreak q_c,\allowbreak k,\allowbreak
d_{\text{soil}},\allowbreak d_p)^\top \in \Theta$, where $\Theta$ represents the physically plausible parameter space. Additional fixed parameters include the typical flow thickness of the erosion zone, maximum observed flow depth, and representative maximum particle diameter $D_p$, all of which are held constant across simulations obtained from \citet{keck2024landslide} (Table~\ref{tab:fixed_params}).

\begin{table}[t]
\caption{Fixed MWR parameters used across all simulations.}
\label{tab:fixed_params}
\centering
\small
\begin{tabular*}{\tblwidth}{@{\extracolsep{\fill}}lcc@{}}
\toprule
\textbf{Parameter} & \textbf{Value} & \textbf{Unit} \\
\midrule
Typical flow thickness (erosion zone) & 4 & m \\
Typical slope (erosion zone) & 0.25 & -- \\
Maximum observed flow depth & 6 & m \\
Maximum particle diameter $D_p$ & 0.316 & m \\
\bottomrule
\end{tabular*}
\end{table}

\textbf{Simulation Procedure.} Given an initial digital elevation model (DEM) $z_0$, a landslide source polygon delineating the failure zone, and the thickness of landslide, the MWR component executes iteratively until all mobilized material has either deposited or exited the model domain. Each iteration $i$ produces an updated elevation field $z_i$, and the complete temporal evolution is captured as:
\begin{equation}
    \{Y_0, Y_1, \ldots, Y_T\} = \mathcal{M}(\theta; z_0, \Omega_{\text{src}}, d_{\text{src}}),
    \label{eq:mwr_simulation}
\end{equation}
where $\mathcal{M}$ denotes the MWR forward operator, $\Omega_{\text{src}}$ is the source polygon, $d_{\text{src}}$ specifies the failure depth, and $T$ is the number of iterations until termination. The output $Y_t = z_t - z_0$ represents the DoD at iteration $t$, with $Y_0 \equiv 0$ corresponding to the pre-event state and $Y_T$ representing the terminal deposited configuration.

\subsection{Data Generation Pipeline}
\label{sec:data_generation}

To assess performance under realistic conditions, we construct an ensemble training paradigm that spans the plausible parameter space while reserving ground truth simulations with unseen parameters for evaluation.

\textbf{Study Site.} We evaluate LAPIS-SHRED on two documented landslide events in the Cascade Mountains near Mount St.\ Helens, Washington, USA. The site experienced a deep-seated failure in 2009 which formed a large-scale debris avalanche over broadly convergent terrain that limited lateral spread of the flow,
followed by a retrogressive enlargement of the same landslide scar and small-scale, channelized debris flow in January 2022.
For the 2009 event, the MWR simulation inputs comprise: (i)~a pre-event DEM at 10-meter native resolution, (ii)~the mapped extent of the 2009 failure as a binary polygon, and (iii)~the failure depth estimated from differencing pre- and post-event DEMs. For the 2022 event, the pre-event DEM corresponds to the post-2009 terrain surface, with the failure extent and depth estimates defined analogously. The two events share similar underlying terrain but differ in failure extent, depth distribution, runout extent, and post-failure topographic boundary conditions, providing complementary test cases for the reconstruction framework.

\textbf{MWR Model Calibration and Runout Scenarios.} For the 2009 scenario experiments, we use a calibrated parameterization of MWR to train LAPIS-SHRED and define sample parameter ranges used in experiments defined below.

\begin{figure*}[t]
\centering
\includegraphics[width=\textwidth]{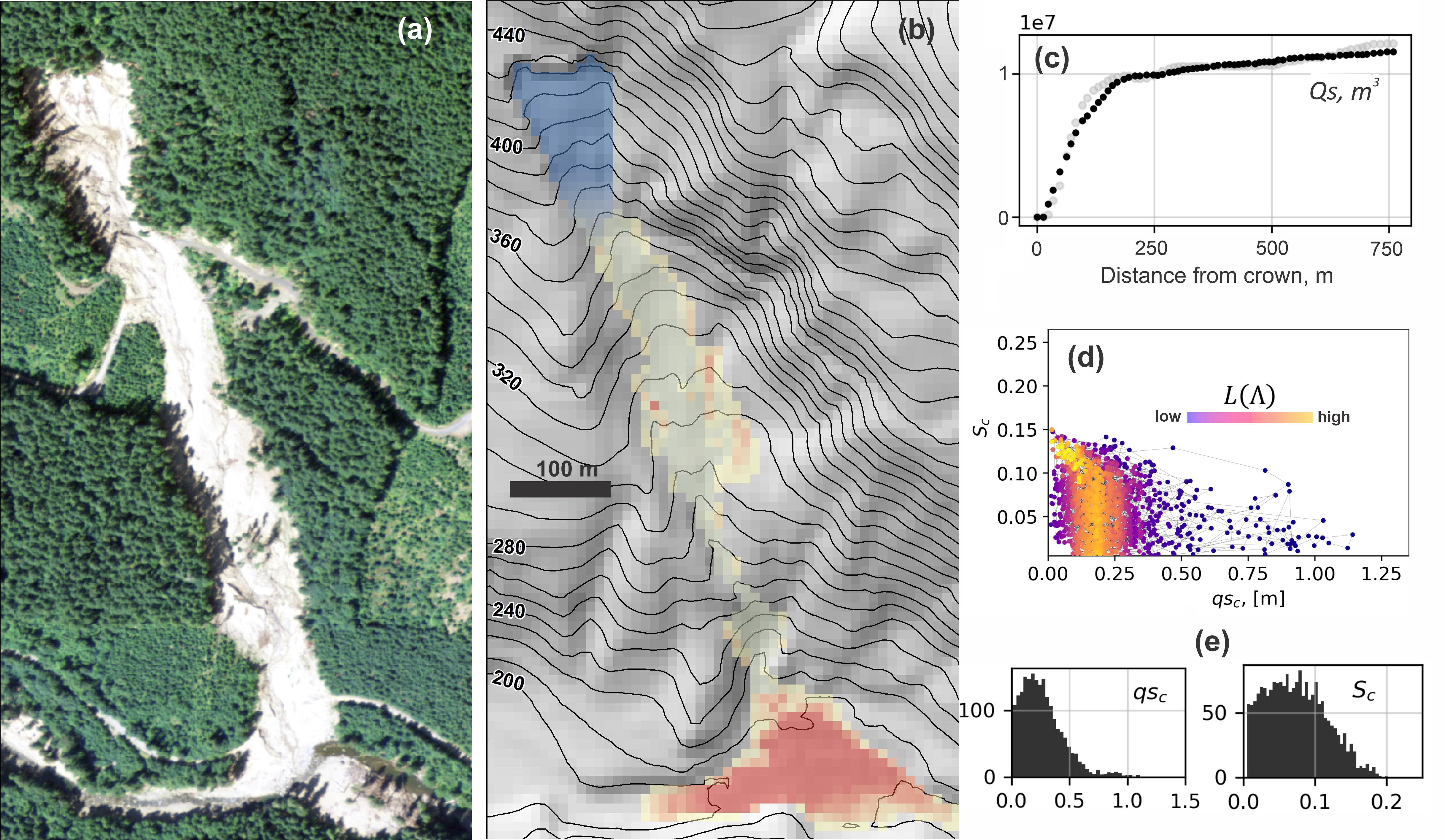}
\caption{(a)~Air photo of the observed Cascade Mountains, 2009 landslide; (b)~Calibrated physics-based modeled runout using MWR. Red indicates deposition and blue indicates erosion relative to the pre-runout terrain; (c)~Calibrated modeled volumetric sediment transport along the runout path relative to observed; (d)~Results of the MCMC calibration algorithm showing the sampled parameter space and likelihood scores; (e)~Histograms of calibration parameters $S_c$ and $q_{sc}$ used to define the calibration (moderate) parameterization used for LAPIS-SHRED model training.}
\label{fig:calibration}
\end{figure*}

We calibrated MWR using a Bayesian Markov Chain Monte Carlo (MCMC) calibration algorithm. The MCMC algorithm iteratively ran MWR over 2000 times. In each iteration, MWR modeled runout was compared to observed runout at a field site and evaluated using a likelihood metric that consisted of a prior probability of the model parameterization, and metrics that reflect how well modeled extent, topographic change and sediment transport matched observed. 
Calibration results for the Cascade Mountains 2009 landslide are shown in \cref{fig:calibration}, where model performance is highest with $S_c > 0.1$ and $q_c < 0.1$. This event was generated by a net water input (rain+snow melt) of ${\sim}200$~mm over a 2-day period in 2009. 
The 2022 landslide was triggered by a similar amount of water input. Considering the wet climate in the region (greater than 3,000 mm/year precipitation) with extremes driven by Atmospheric River events, we classify this event as moderately wet for LAPIS-SHRED training purposes. Additional details on the calibration algorithm are detailed in \citet{keck2024landslide}. 

Landslide runout reflects a broad spectrum of mechanical behavior governed by sediment properties, water content, topography, lithologic setting, and basal resistance. At one end, relatively dry granular runout or grain-dominated debris-flow behavior is characterized by flow mechanics dominated by interactions among solid particles (sand, gravel, and boulders). In such cases, grain contacts, basal resistance, and internal friction exert primary control on mobility, typically producing shorter runout and a stronger tendency for deposition. At the other end, water-laden debris flows (10-40 percent solid concentration) or mud-rich slurry-like flows (>40 percent mud concentration) can exhibit substantially greater mobility, longer travel distances, and higher bed scour or sediment entrainment due to the influence of the sediment–water mixture. This variability is commonly interpreted as a continuum from frictional, grain-dominated motion to more fluid-rich debris-flow behavior, rather than as a set of discrete classes \citep{iverson1997physics,hungr2014varnes}. Empirical observations in \citet{keck2024landslide} indicate dry, coarse-grained landslides (e.g., the Rocky Mountains Landslide) tend to stop on higher-slopes while wetter, finer-grained landslides deposit on lower slopes (e.g., the Black Hills South landslide).

In the present study, this continuum is represented using prescribed parameter sets within MWR, rather than by explicitly resolving the full mechanics of debris-flow rheology. We created a high-mobility or wet, fine-grained, slurry-like flow response with longer runout and greater potential for scour or entrainment; and a lower-mobility (grain-dominated) response with limited runout and rapid deposition, by adjusting the critical slope and critical flux parameters. In this conceptualization, the calibrated parameterization represents a moderate-mobility scenario between the two. These cases are used as empirical mobility scenarios, not as fixed rheological classes or direct proxies for measured water content. Such wide range of behavior enables us to explore the capabilities of the LAPIS-SHRED methodology.

\textbf{Ensemble Generation.} For each experimental scenario, we generate an ensemble of $K = 15$ training simulations by sampling each component of $\boldsymbol{\theta}$ independently and uniformly from parameter ranges representative of dry, moderate, and wet landslide conditions informed by the 2009 calibration and the mobility scenarios described above:
\begin{equation}
    \theta_j^{(k)} \sim \mathcal{U}(\theta_j^{\min}, \theta_j^{\max}), \quad j = 1, \ldots, 5, \quad k = 1, \ldots, K,
    \label{eq:bounded_sampling}
\end{equation}
where the bounds $[\theta_j^{\min}, \theta_j^{\max}]$ are specified per scenario based on domain expertise.

We organize experiments into scenarios defined by climate class (dry, moderate, wet) and soil availability, producing parameter regimes that span the range from low-mobility earth flows (high $S_c$, high $q_c$, low $k$) to high-mobility debris flows (low $S_c$, low $q_c$, high $k$). We additionally evaluate four ``general site model'' scenarios with broader parameter ranges. In total, 16 scenarios are defined; full parameter specifications are provided in \cref{app:scenarios}.

For each parameter vector, $\theta^{(k)}$, we execute the MWR model to obtain complete spatiotemporal trajectory:
\begin{equation}
    \mathcal{D}^{(k)} = \{Y^{(k)}_0, Y^{(k)}_1, \ldots, Y^{(k)}_{T_k}\} = \mathcal{M}(\theta^{(k)}),
    \label{eq:training_data}
\end{equation}
where $T_k$ denotes the number of iterations for trajectory $k$. The training dataset aggregates all ensemble trajectories: $\mathcal{D}_{\text{train}} = \bigcup_{k=1}^{K} \mathcal{D}^{(k)}$.

\textbf{Ground Truth Test Data.} For each scenario, a ground truth parameter vector based on MWR model outputs $\boldsymbol{\theta}_{\text{true}}$ is defined at the boundary of the parameter ranges, lying within the convex hull of the training ensemble but not identical to any training configuration. The ground truth simulation:
\begin{equation}
    \mathcal{D}_{\text{test}} = \{Y^{\text{obs}}_0, Y^{\text{obs}}_1, \ldots, Y^{\text{obs}}_{T}\} = \mathcal{M}(\theta_{\text{true}})
    \label{eq:test_data}
\end{equation}
serves as the evaluation target. Critically, during inference, only a short observation window is provided to the model; the remainder of the trajectory is reserved for post-hoc evaluation.

\textbf{Preprocessing.} Two preprocessing steps are applied before training. First, trailing simulation frames where the active boundary has ceased expanding are trimmed (\emph{stagnation cropping}), removing physically uninformative tail frames where the MWR model continues iterating after flow cessation. Second, training simulations whose duration deviates by more than a factor of 2.5 from the ground truth length are excluded, ensuring the ensemble provides temporally comparable trajectories. Details are reported in \cref{app:preprocessing}.

\textbf{Sparse Sensor Configuration.} To simulate field observation constraints, we employ $p = 32$ virtual sensors distributed across the spatial domain. Approximately 60\% (18 sensors) are placed within the active region $\mathcal{A} = \{(i, j) : \exists\, t,\, |Y_t(i,j)| > \tau\}$ with probability proportional to the maximum DoD magnitude at each cell, ensuring dense sampling where geophysical activity concentrates; the remaining 40\% (14 sensors) are distributed across quiescent regions, providing static reference points for reconstruction stability. This \emph{default} DoD-weighted placement yields a sampling ratio of $p/n \approx 0.33\%$. We additionally evaluate a \emph{top-clustering} configuration, where all active-region sensors are concentrated at the landslide initiation zone, simulating deployment constraints (\cref{sec:sensor_placement}).

\section{LAPIS-SHRED for Landslide Reconstruction}
\label{sec:method}

We apply LAPIS-SHRED~\citep{bao2026latent} to reconstruct the complete spatiotemporal evolution of landslide runout events from sparse sensor observations confined to short temporal windows. This section formulates the inference problem, summarizes the LAPIS-SHRED framework and the components relevant to our landslide application, and describes the training and inference protocols.
We additionally introduce SINDy-SHRED~\citep{gao2026sparse} as a baseline at the end of this section 
to answer the natural question of whether, given denser observations, the SHRED framework could identify interpretable governing equations for the latent dynamics; empirical comparisons are presented in \cref{sec:baselines}.

\subsection{Problem Formulation}
\label{sec:problem}

Let $\mathbf{Y}(t) \in \mathbb{R}^{H \times W}$ denote the full spatial state at time $t$, representing the difference-of-DEMs (DoD) field that characterizes landslide evolution. A physics-based simulator $\mathcal{M}$ parameterized by $\boldsymbol{\theta} = (S_c, q_c, k, d_{\text{soil}}, d_{\text{param}})^\top$ generates the temporal evolution:
\begin{equation}
\{\mathbf{Y}_0, \mathbf{Y}_1, \ldots, \mathbf{Y}_T\} = \mathcal{M}(\boldsymbol{\theta}),
\label{eq:simulation}
\end{equation}
where $\mathbf{Y}_0 \equiv \mathbf{0}$ represents the initial pre-event state and $\mathbf{Y}_T$ is the terminal deposited configuration.

Sparse sensor observations are obtained via a sampling operator $\mathbf{M} \in \mathbb{R}^{p \times n}$ with $p \ll n = H \times W$:
\begin{equation}
\mathbf{s}_t = \mathbf{M} \, \text{vec}(\mathbf{Y}_t) \in \mathbb{R}^p.
\label{eq:sensor_sampling}
\end{equation}
We evaluate two complementary inference settings:

\emph{Backward inference (terminal-only observation).} The model receives only the terminal sensor observation $\mathbf{s}_T^{\text{obs}} = \mathbf{M} \, \text{vec}(\mathbf{Y}_T^{\text{obs}})$, representing the post-event topographic change (erosion and deposition). The task is to reconstruct the complete antecedent evolution $\{\hat{\mathbf{Y}}_0, \ldots, \hat{\mathbf{Y}}_T\}$. This setting reflects applications in which only post-event observations (e.g., field surveys, DoD, or remote sensing products) are available. It also enables conditional trajectory reconstruction given a specified terminal state, which may support retrospective analysis and reconstruction of past debris flows.

\emph{Forward inference (short initial window).} The model receives the first $W$ frames of sensor observations $\{\mathbf{s}_0^{\text{obs}}, \ldots, \allowbreak
\mathbf{s}_W^{\text{obs}}\}$, where $W \ll T$ represents a short initial observation window (e.g., 10--20\% of the trajectory). The task is to predict the remaining evolution $\{\hat{\mathbf{Y}}_{W+1}, \ldots, \hat{\mathbf{Y}}_T\}$. This setting represents cases where partial early-stage observations are available and the objective is to forecast downstream dynamics. Forward inference is ideal for regions where unstable terrain is monitored for early warning purposes for slow-moving landslides and long runout events. It also allows projection of trajectories from prescribed initial configurations, supporting prospective analysis under assumed initial conditions.

In both modes, the unobserved ground truth frames are never provided to any model component during inference and are reserved exclusively for post-hoc evaluation. The sequence length $T$ is assumed known from event duration metadata, which is typically available from seismic records or remote sensing in field applications.

\begin{remark}[Extreme Data Constraint]
\label{rem:data_efficiency}
During inference, \textbf{only sparse sensor measurements} from the observed window are available---not full spatial fields at the observed frames, and not any intermediate states. In the backward setting, this reduces to $p \ll n$ measurements at a single terminal timepoint. This represents an extreme low-data regime: reconstructing $T$ full spatiotemporal frames from a short observation window of hyper-sparse sensor measurements.
\end{remark}

\subsection{LAPIS-SHRED for Landslide Reconstruction}
\label{sec:framework}

LAPIS-SHRED is a modular architecture for reconstructing or forecasting complete spatiotemporal dynamics from sparse sensor measurements confined to short temporal windows. It operates through a three-stage pipeline, summarized here; see \citet{bao2026latent} for the full architectural formulation and analysis.

\textbf{Stage~1: SHRED module.} A Seq2Seq~\citep{sutskever2014sequence} formulation of the SHRED architecture~\citep{williams2024sensing} is trained on simulation ensembles to map sparse sensor time-histories to a structured latent space. The bidirectional LSTM~\citep{hochreiter1997long} processes the full sensor sequence and produces a latent trajectory $\mathbf{Z} = \{\mathbf{z}_0, \ldots, \mathbf{z}_T\}$, which a shallow decoder maps to the full spatiotemporal state. Training employs a weighted reconstruction loss that emphasizes the active region $\mathcal{A}$:
\begin{equation}
\mathcal{L}_{\text{SHRED}} = \frac{1}{T} \sum_{t=1}^{T} \sum_{i=1}^{n} w_i \left( \hat{Y}_{t,i} - Y_{t,i} \right)^2,
\label{eq:shred_loss}
\end{equation}
where $w_i = \alpha$ for cells in the active region $\mathcal{A} = \{i : \exists\, t,\, |Y_{t,i}| > \tau\}$ and $w_i = 1$ otherwise. The elevated weight $\alpha$ focuses reconstruction accuracy on the geophysically relevant domain, which constitutes less than 5\% of the spatial grid.
For backward inference, the SHRED module is trained with static padding
(replicating a single observation into a fixed-length input sequence):
$L$ copies of the terminal sensor vector are appended to each training sequence, conditioning the temporal unit to recognize constant input signals as indicative of system quiescence. At inference, the terminal observation is replicated into a pseudo-sequence of length $L$, enabling latent encoding from a single frame. This strategy exploits the terminal stationarity of the deposited landslide state ($\mathbf{Y}_{T+\delta} \approx \mathbf{Y}_T$ for $\delta > 0$).

\textbf{Stage~2: Temporal model.} A Seq2Seq temporal dynamics model, trained on simulation-derived latent trajectories, learns to propagate latent states forward or backward in time from the observed window to span unobserved temporal regions. The temporal model produces the complete unobserved latent trajectory in a single forward pass, avoiding the error accumulation inherent in autoregressive rollout. Training uses a composite objective combining reconstruction and structural (trajectory-shape) consistency terms; implementation details are provided in~\cref{app:implementation}.

\textbf{Stage~3: Inference.} At deployment, only the short observation window from the ground truth system is provided. The frozen SHRED temporal unit extracts latent codes from the observed window, the temporal model infers the unobserved portion of the latent trajectory, and the frozen SHRED decoder reconstructs the complete spatiotemporal evolution. The SHRED components remain fixed during temporal model training and inference, enabling a modular formulation. 
Inference from a single observation is made possible by two complementary learned mappings: SHRED provides a mapping from sparse sensor measurements to a compact latent representation of the full spatial state, while the temporal model---trained on the ensemble of simulation-derived latent trajectories---provides a mapping from the latent state at observed timepoints to the latent states at all other times. Composing these mappings bridges the gap from sparse sensor readings at one timepoint to the complete spatiotemporal field.
Figure~\ref{fig:lapis_architecture} illustrates the pipeline.

\textbf{Terminal Stationarity.} A critical physical property exploited by LAPIS-SHRED is the \emph{terminal stationarity} of the final state: once all the landslide runout material has deposited and flow has ceased, the DoD field stabilizes such that $Y_{T+\delta} \approx Y_T$ for $\delta > 0$. This property enables the static terminal padding strategy for backward inference.

\textbf{Variable-length sequence handling.} The simulation ensembles generated under different MWR parameter configurations exhibit variable sequence lengths $T_k$. To accommodate this heterogeneity during training, we grouped sequences into length-based buckets with masked losses, avoiding the computational overhead of global padding while preserving training efficiency across heterogeneous sequence lengths.

\subsection{Training and Inference Protocols}
\label{sec:training_inference}

\textbf{Training.} For each scenario, the Seq2Seq SHRED is trained on the $K$ ensemble trajectories $\mathcal{D}_{\text{train}}$, with scalers fitted on the combined dataset. Latent trajectories are then extracted from the frozen SHRED temporal unit, and the temporal model is trained on these latent sequences using the bucketed procedure described above. Both the SHRED model and the temporal model are trained exclusively on simulation data; no ground truth observations enter the training pipeline. Full data separation and evaluation protocols are detailed in~\cref{app:data_separation}; architecture specifications and hyperparameters are reported in~\cref{app:implementation}.

\textbf{Backward inference.} Given only the terminal sensor observations $\mathbf{s}_T^{\text{obs}}$ from a potentially unseen parameter configuration $\boldsymbol{\theta}^{\text{test}} \notin \Theta_{\text{train}}$, reconstruction proceeds as: (1)~construct a static-padded pseudo-sequence of $L$ copies of $\mathbf{s}_T^{\text{obs}}$; (2)~encode via the frozen SHRED temporal unit to obtain the terminal latent $\mathbf{z}_T^{\text{obs}}$; (3)~generate $\{\hat{\mathbf{z}}_0, \ldots, \hat{\mathbf{z}}_{T-1}\} = \mathcal{B}(\mathbf{z}_T^{\text{obs}}, T; \psi^*)$ via the backward temporal model; (4)~form the complete latent trajectory $\hat{\mathbf{Z}} = [\hat{\mathbf{z}}_0, \ldots, \hat{\mathbf{z}}_{T-1}, \mathbf{z}_T^{\text{obs}}]$; (5)~decode via the frozen SHRED decoder.

\textbf{Forward inference.} Given the initial observation window $\{\mathbf{s}_0^{\text{obs}}, \ldots, \mathbf{s}_W^{\text{obs}}\}$, prediction proceeds as: (1)~encode the observed frames via the frozen SHRED temporal unit to obtain $\{\mathbf{z}_0^{\text{obs}}, \ldots, \mathbf{z}_W^{\text{obs}}\}$; (2)~generate $\{\hat{\mathbf{z}}_{W+1}, \ldots, \hat{\mathbf{z}}_T\} = \mathcal{F}(\{\mathbf{z}_0^{\text{obs}}, \ldots, \mathbf{z}_W^{\text{obs}}\}, T; \psi^*)$ via the forward temporal model; (3)~form the complete latent trajectory and decode.

The entire pipeline operates in a single forward pass, requiring no iterative optimization or access to ground truth beyond the observed window sensor measurements.

\subsection{SINDy-SHRED Baseline}
\label{sec:sindy_method}

SINDy-SHRED~\citep{gao2026sparse} augments a frame-by-frame SHRED framework with Sparse Identification of Nonlinear Dynamics (SINDy)~\citep{brunton2016discovering} regularization, discovering latent governing equations of the form $\dot{\mathbf{z}} = \boldsymbol{\Xi}^\top \boldsymbol{\Theta}(\mathbf{z})$, where $\boldsymbol{\Theta}$ is a library of candidate functions and $\boldsymbol{\Xi}$ a sparse coefficient matrix co-trained with the network and refined via post-hoc sparse regression. Because the frame-by-frame temporal unit requires sensor observations at every timestep, SINDy-SHRED operates in a dense-temporal-observation regime. Forward prediction integrates the discovered equations from an observed latent state, yielding interpretable dynamics but incurring error accumulation over longer horizons. An empirical comparison of SINDy-SHRED and LAPIS-SHRED on the landslide reconstruction tasks is presented in \cref{sec:baselines}.

\section{Results}
\label{sec:results}

We evaluate LAPIS-SHRED on landslide runout reconstructions near Mount St.\ Helens, assessing the capacity to recover complete spatiotemporal dynamics under two inference modes: backward inference from a single terminal observation, and forward inference from a short initial observation window. We consider two landslide events at the site---the 2009 deep-seated failure and the 2022 retrogressive enlargement---spanning distinct runout regimes and parameter configurations. All experiments employ the Seq2Seq SHRED backbone trained on an ensemble of $K = 15$ MWR simulations with $p = 32$ sensors, as described in \cref{sec:data_generation,sec:framework}.

We report four primary metrics throughout. The normalized root mean squared error (NRMSE), defined as the active-region RMSE divided by the ground truth field range $\Delta = \max - \min$, provides a scale-invariant measure of reconstruction fidelity. The Structural Similarity Index Measure (SSIM)~\citep{wang2004image} captures preservation of spatial patterns and runout morphology. The Nash--Sutcliffe efficiency (NSE)~\citep{nash1970river}, standard in hydrological and geomorphological modeling, quantifies the fraction of variance explained by the reconstruction relative to the observed mean. The modified Kling--Gupta efficiency (KGE$'$)~\citep{gupta2009decomposition} provides a balanced assessment of correlation, variability bias, and mean bias. For the 2022 event, we additionally report active-region RMSE in physical units (meters), following the conventions of the earlier results; for the SINDy-SHRED comparison in Section~\ref{sec:baselines} and field validation in Section~\ref{sec:observed_validation}, Pearson correlation is also reported for spatial agreement.
RMSE and NRMSE values are computed over the active region $\mathcal{A}$ unless otherwise specified, as reconstruction accuracy in the geophysically relevant domain is the primary concern. For each configuration, we report both the LAPIS-SHRED reconstruction and the corresponding SHRED reconstruction with access to the full sensor time-series, which serves as an upper bound on achievable performance under complete temporal observation.

\subsection{LAPIS-SHRED Reconstruction Performance}
\label{sec:results_core}

\subsubsection{2009 Event: Backward and Forward Inference Across Scenarios}
\label{sec:results_2009_scenario}

We evaluate LAPIS-SHRED on four representative scenarios derived from the 2009 event, spanning distinct climate and soil conditions; each scenario is named as \emph{climate class}--\emph{moisture subclass}--\emph{soil availability}, reflecting the three axes of the parameter matrix (\cref{tab:scenario_params}): 
(i)~Dry--Dryer--Shallow\,Soil, a shallow failure under dry conditions with limited soil availability; (ii)~Moderate--Wetter--Deep\,Soil, a moderate-depth failure under wetter conditions with high soil availability; (iii)~Wet--Dryer--Deep\,Soil, a deep, fast-propagating flow under wet conditions; and (iv)~General--Low\,$S_c$--Low\,$k$, a broadly parameterized scenario with low critical slope and low erosion coefficient. Each scenario employs $K = 15$ ensemble training simulations with ground truth parameters lying within the convex hull of the training ensemble but not included in training. Per-scenario reconstruction figures are provided in \cref{app:scenario_figures}.

For each scenario, we evaluate both inference modes. In backward inference, the model receives only the terminal sensor observation $\mathbf{s}_T^{\text{obs}}$. In forward inference, the model receives the first $W$ frames of sensor observations, where $W = \max(5, \lfloor 0.10 \cdot T_{\text{gt}} \rfloor)$, corresponding to approximately 10\% of the ground truth trajectory.

\Cref{tab:lapis_2009} summarizes the results. Across all four scenarios and both inference modes, LAPIS-SHRED achieves NRMSE below 5.8\% and SSIM above 0.961, closely approaching the SHRED upper bound that has access to the complete temporal record. Several patterns emerge from the comparison.

\textbf{Overall reconstruction performance.} LAPIS-SHRED achieves consistently low reconstruction error across the four representative scenarios, with active-region NRMSE ranging from 3.8\% to 5.8\% and SSIM remaining high across all configurations, indicating that the latent temporal model can recover or predict full runout evolution from highly restricted temporal observations.

\textbf{Forward and backward inference are broadly comparable.} Across the four scenarios, the two inference modes yield similar reconstruction quality, with NRMSE differences generally within approximately one percentage. This suggests that neither the terminal deposit nor the initial observation window is uniformly more informative; instead, the relative conditioning is governed by the runout regime and the temporal distribution of geophysically relevant dynamics. Nevertheless, backward inference exhibits higher variance across random seeds than forward inference, consistent with the reduced conditioning information available from a single terminal frame compared to a multi-frame observation window.

\textbf{Comparison with the SHRED upper bound.} The full-observation SHRED upper bound achieves NRMSE of 0.033--0.044, confirming that the Seq2Seq SHRED accurately represents the simulated spatiotemporal dynamics when complete sensor histories are available. LAPIS-SHRED remains close to this upper bound in most configurations, with NRMSE gaps below 1.6\%, demonstrating that LAPIS-SHRED effectively recovers spatiotemporal dynamics under severe observational constraints.

\begin{table*}[t]
\centering
\caption{LAPIS-SHRED reconstruction performance on four representative 2009 Mount St.\ Helens scenarios. For each scenario, backward inference uses only the terminal observation $\mathbf{s}_T$; forward inference uses the first $W$ frames (approximately 10\% of the trajectory). The SHRED upper bound uses the full sensor time-series. All NRMSE values are computed over the active region; SSIM over the full domain.}
\label{tab:lapis_2009}
\small
\setlength{\tabcolsep}{4pt}
\resizebox{\tblwidth}{!}{%
\begin{tabular}{ll cc cc cc}
\toprule
 & & \multicolumn{2}{c}{\textbf{SHRED (upper bound)}} & \multicolumn{4}{c}{\textbf{LAPIS-SHRED}} \\
\cmidrule(lr){3-4} \cmidrule(lr){5-8}
\textbf{Scenario} & \textbf{Mode} & \textbf{NRMSE} & \textbf{SSIM} & \textbf{NRMSE} & \textbf{SSIM} & \textbf{NSE} & \textbf{KGE$'_{\text{abs}}$} \\
\midrule
  Dry--Dryer--Shallow Soil & Bwd ($\mathbf{s}_T$) & $0.042 \pm 0.001$ & $0.987 \pm 0.002$ & $0.057 \pm 0.014$ & $0.977 \pm 0.008$ & $0.806 \pm 0.090$ & $0.795 \pm 0.059$ \\
    & Fwd ($W=5$) & $0.042 \pm 0.001$ & $0.988 \pm 0.002$ & $0.049 \pm 0.006$ & $0.984 \pm 0.003$ & $0.859 \pm 0.029$ & $0.847 \pm 0.018$ \\
  \addlinespace
  Moderate--Wetter--Deep Soil & Bwd ($\mathbf{s}_T$) & $0.034 \pm 0.001$ & $0.982 \pm 0.002$ & $0.048 \pm 0.009$ & $0.972 \pm 0.007$ & $0.705 \pm 0.111$ & $0.690 \pm 0.096$ \\
    & Fwd ($W=10$) & $0.036 \pm 0.002$ & $0.981 \pm 0.002$ & $0.038 \pm 0.001$ & $0.978 \pm 0.001$ & $0.827 \pm 0.007$ & $0.812 \pm 0.010$ \\
  \addlinespace
  Wet--Dryer--Deep Soil & Bwd ($\mathbf{s}_T$) & $0.033 \pm 0.004$ & $0.984 \pm 0.003$ & $0.048 \pm 0.009$ & $0.973 \pm 0.005$ & $0.697 \pm 0.113$ & $0.687 \pm 0.084$ \\
    & Fwd ($W=9$) & $0.033 \pm 0.003$ & $0.983 \pm 0.003$ & $0.042 \pm 0.006$ & $0.976 \pm 0.005$ & $0.766 \pm 0.067$ & $0.736 \pm 0.054$ \\
  \addlinespace
  General--Low $S_c$--Low $k$ & Bwd ($\mathbf{s}_T$) & $0.044 \pm 0.002$ & $0.972 \pm 0.001$ & $0.058 \pm 0.010$ & $0.961 \pm 0.008$ & $0.499 \pm 0.171$ & $0.515 \pm 0.108$ \\
    & Fwd ($W=9$) & $0.044 \pm 0.001$ & $0.972 \pm 0.001$ & $0.043 \pm 0.001$ & $0.975 \pm 0.002$ & $0.751 \pm 0.019$ & $0.709 \pm 0.020$ \\
\bottomrule
\end{tabular}%
}
\end{table*}

\begin{figure}[pos=thbp]
\centering
\begin{subfigure}[t]{\columnwidth}
    \centering
    \includegraphics[width=\columnwidth]{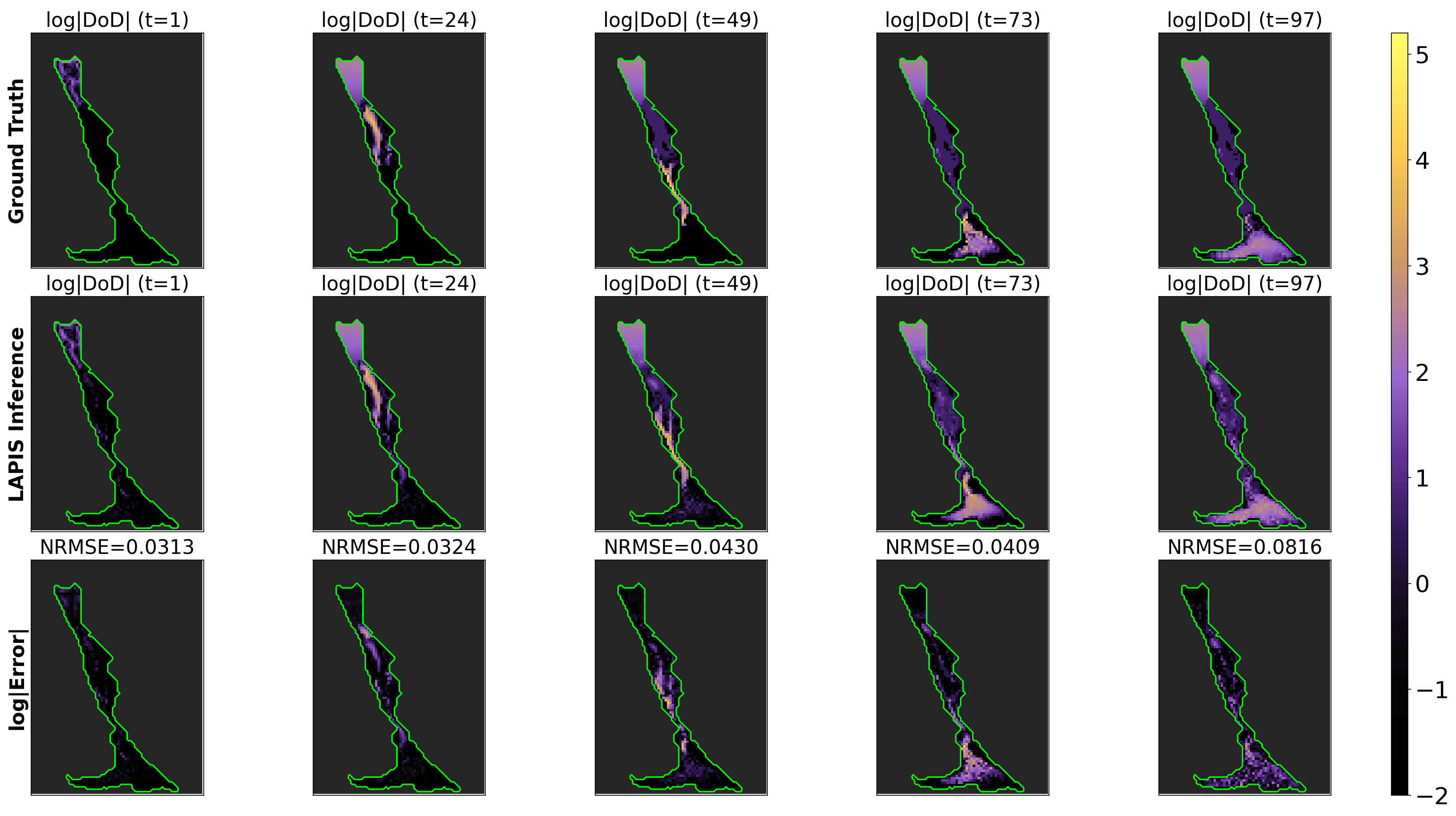}
    \caption{Backward inference from terminal observation.}
    \label{fig:lapis_2009_backward}
\end{subfigure}
\\[6pt]
\begin{subfigure}[t]{\columnwidth}
    \centering
    \includegraphics[width=\columnwidth]{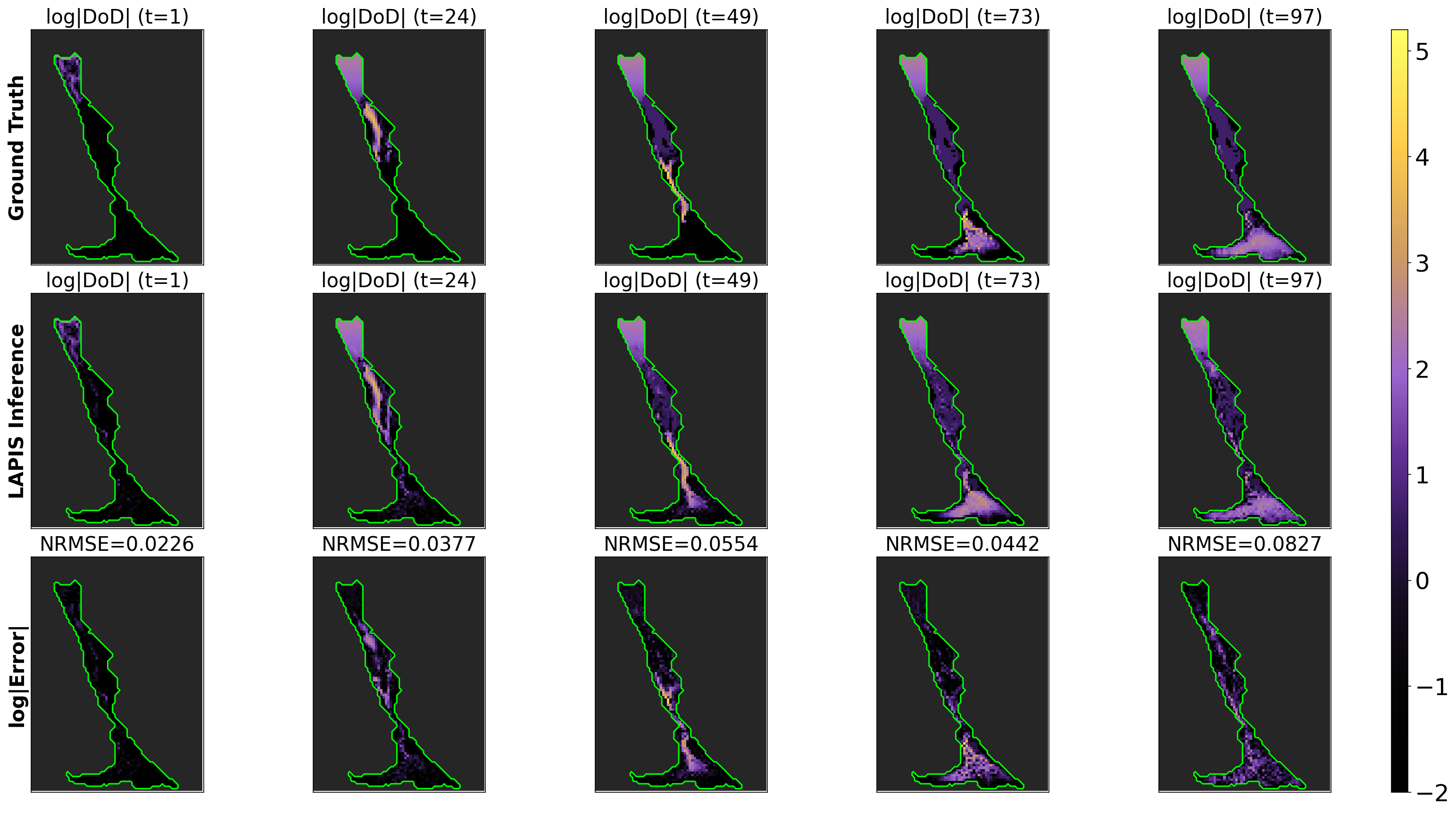}
    \caption{Forward inference from initial observation window ($W = 9$ frames).}
    \label{fig:lapis_2009_forward}
\end{subfigure}
\caption{LAPIS-SHRED reconstruction of the 2009 Mount St.\ Helens event under the Wet--Dryer--Deep\,Soil scenario. Each panel shows: ground truth DoD (top row), LAPIS-SHRED reconstruction (middle row), and pointwise error with per-frame RMSE (bottom row). Columns correspond to selected timesteps. Log-scale visualization enhances contrast for small-magnitude changes; green contours outline the active region boundary.}
\label{fig:lapis_2009_example}
\end{figure}

\Cref{fig:lapis_2009_example} presents the LAPIS-SHRED reconstruction for the Wet--Dryer--Deep\,Soil scenario under both inference modes. The backward reconstruction (top) faithfully recovers the spatial progression, with errors primarily concentrated in regions of large, transient DoD magnitude along the active runout path. The forward prediction (bottom), starting from only the first 9 observed frames (approximately 10\% of the trajectory), also captures the overall runout extent and deposition pattern. Both modes preserve the overall spatial structure of the landslide process, as reflected in SSIM values of 0.973 (backward) and 0.976 (forward).

\subsubsection{2022 Event: Backward Inference Validation}
\label{sec:results_2022_backward}

In addition to the multi-scenario 2009 event analysis, we also evaluate LAPIS-SHRED on the 2022 retrogressive failure at the same site, which formed a much smaller and more channelized debris flow, under a single parameter regime with backward inference. 
Unlike the 2009 scenario experiments in \cref{sec:results_2009_scenario}, which are organized by climate class, here we use a single training ensemble spanning broad parameter ranges not tied to a specific runout regime.
The simulated ground truth system---a held-out MWR simulation---is governed by parameters $\theta_{\text{true}} = (S_c = 0.21,\, q_c = 0.40,\, k = 0.045)^\top$, which lies within the convex hull of the training ensemble but was not included in training. Sensors are distributed using the default DoD-weighted placement strategy described in \cref{sec:data_generation};
sparse sensing is maintained to evaluate the method under potential monitoring constraints, where full spatial coverage may not be available within the time window required for rapid response.

\Cref{tab:lapis_2021} presents the reconstruction metrics. The SHRED upper bound, which has access to the complete sensor time-history $\{\mathbf{s}_1, \ldots, \mathbf{s}_T\}$, achieves an active-region RMSE of 0.867~m and SSIM of 0.981. LAPIS-SHRED, operating under the extreme constraint of observing only the terminal sensor measurements $\mathbf{s}_T^{\text{obs}}$, achieves RMSE of 1.065~m, NRMSE of 0.034, and SSIM of 0.979, indicating faithful recovery of the runout evolution from a single terminal snapshot.

\begin{table}[t]
\centering
\caption{LAPIS-SHRED backward inference on the 2022 Mount St.\ Helens event (default sensor configuration). SHRED uses the full sensor time-series; LAPIS-SHRED uses only the terminal observation. Metrics computed over the active region except SSIM (full domain).}
\label{tab:lapis_2021}
\small
\resizebox{\tblwidth}{!}{%
\begin{tabular}{lccccc}
\toprule
\textbf{Method} & \textbf{Input} & \textbf{RMSE (m)} & \textbf{NRMSE} & \textbf{SSIM} \\
\midrule
SHRED (upper bound) & $\{\mathbf{s}_1, \ldots, \mathbf{s}_T\}$ & 0.867 & 0.028 & 0.981 \\
LAPIS-SHRED         & $\mathbf{s}_T$ only                       & 1.065 & 0.034 & 0.979 \\
\bottomrule
\end{tabular}%
}
\end{table}

\begin{figure}[pos=thbp]
\centering
\includegraphics[width=\columnwidth]{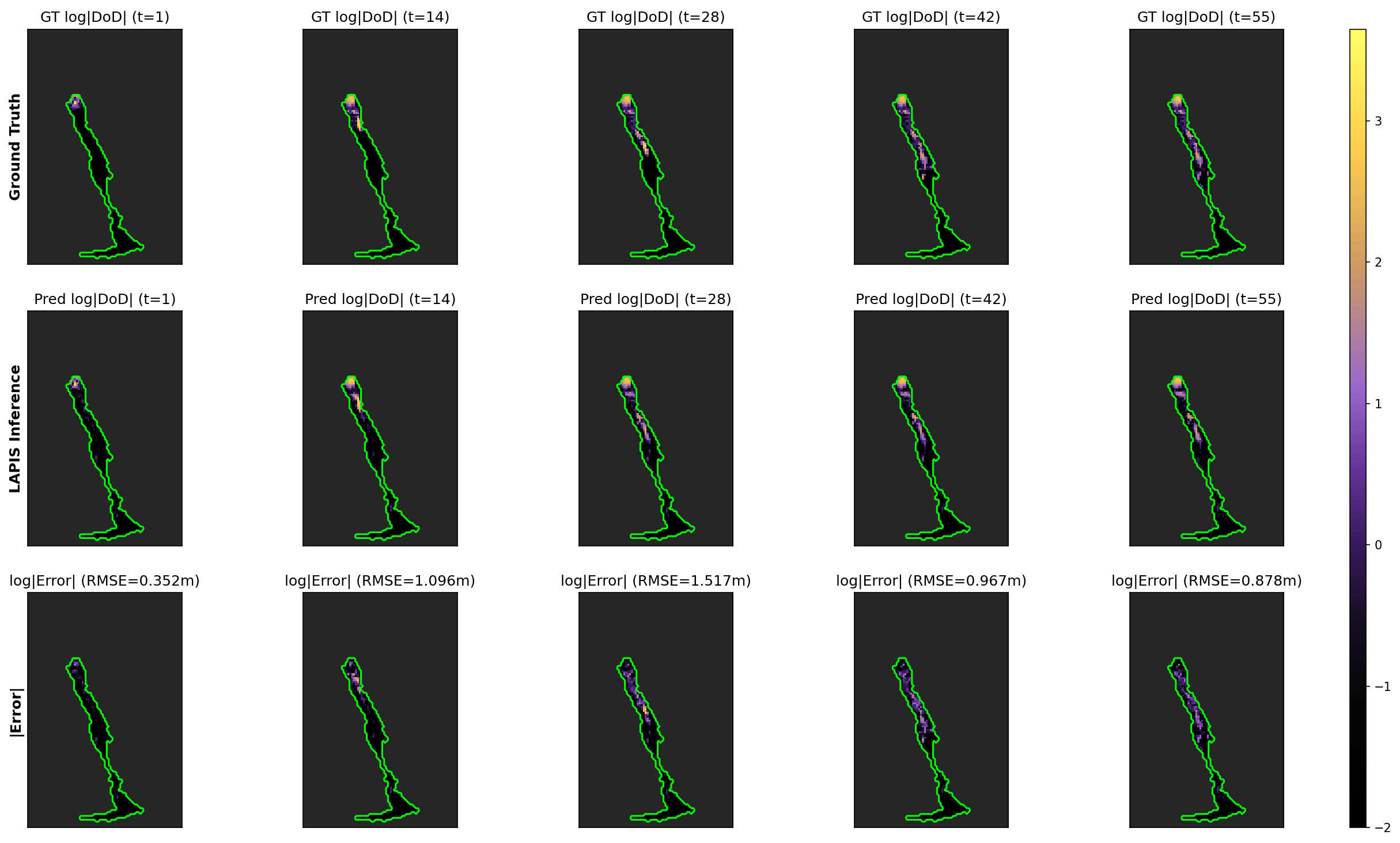}
\caption{LAPIS-SHRED backward reconstruction of the 2022 Mount St.\ Helens event using the default sensor placement. Rows: ground truth DoD (top), LAPIS-SHRED reconstruction (middle), and pointwise error with per-frame RMSE (bottom). Columns correspond to selected timesteps. Log-scale visualization enhances contrast for small-magnitude changes; green contours outline the active region boundary.}
\label{fig:lapis_2021_backward}
\end{figure}

\Cref{fig:lapis_2021_backward} presents the spatiotemporal reconstruction. The predicted DoD field closely tracks the ground truth across the full temporal evolution, with errors concentrated along the main runout path where the largest elevation changes occur. The reconstruction captures both the spatial extent of the failure zone and the downstream deposition pattern, demonstrating that the backward model successfully recovers the antecedent dynamics from the terminal deposited state.

\subsection{Baseline Comparisons}
\label{sec:baselines}

We evaluate LAPIS-SHRED against two classes of baselines that contextualize its performance from complementary perspectives: simulation-interpolation methods that operate under the same terminal-only observation constraint, and an equation-discovery approach (SINDy-SHRED) that achieves competitive reconstruction but requires substantially denser temporal observations.
These comparisons address two questions. First, can non-learned methods such as simulation interpolation achieve comparable quality under the same observation constraints? Second, how does our framework compare with a SINDy-based latent approach that yields interpretable latent dynamics but operates under greater observation requirements?

\subsubsection{Simple Interpolation Baselines (2022 Event)}

To assess whether a learned latent model is necessary for trajectory reconstruction, we compare LAPIS-SHRED against four simple baselines that reconstruct spatiotemporal trajectories by combining the $K$ training simulations using only the ground truth terminal sensor observations $\mathbf{s}_T^{\text{obs}}$---the same information available to LAPIS-SHRED at inference. All baselines first temporally align the training trajectories to the ground truth length via linear interpolation, then form a weighted combination: (i)~\emph{Simple Average}, the unweighted mean of all $K$ training trajectories; (ii)~\emph{Sensor-Weighted Average}, a Gaussian-kernel weighting $w_k \propto \exp(-\|\mathbf{s}_T^{(k)} - \mathbf{s}_T^{\text{obs}}\|^2 / 2\sigma^2)$ based on terminal sensor mismatch; (iii)~\emph{Nearest Neighbor}, which selects the single training trajectory with the closest terminal sensor match; and (iv)~\emph{Ridge Interpolation}, which solves a ridge-regularized least-squares problem for the optimal linear combination weights.

\Cref{tab:baselines_interpolation} reports the results on the 2022 Mount St.\ Helens ground truth with the default sensor configuration. All four baselines yield active-region RMSE values in the range of 1.68--2.01~m (NRMSE 0.046--0.055), substantially higher than the LAPIS-SHRED RMSE of 1.065~m (NRMSE 0.034). The best-performing baseline (Sensor-Weighted Average) exceeds the LAPIS-SHRED error by 57\% in RMSE and 35\% in NRMSE, confirming that the learned latent model with backward inference provides a substantial advantage over direct interpolation from the simulation ensemble.

\begin{table}[t]
\caption{Simple interpolation baselines on the 2022 Mount St.\ Helens ground truth ($K = 15$ training simulations, $\pm 100\%$ perturbation, default sensor configuration). All methods use only terminal sensor observations. LAPIS-SHRED result from \cref{tab:lapis_2021} shown for reference.}
\label{tab:baselines_interpolation}
\centering
\begin{tabular*}{\tblwidth}{@{\extracolsep{\fill}}l cc@{}}
\toprule
\textbf{Method} & \textbf{RMSE (m)} & \textbf{NRMSE} \\
\midrule
Simple Average           & 1.881 & 0.054 \\
Sensor-Weighted Average  & 1.676 & 0.046 \\
Nearest Neighbor         & 2.005 & 0.055 \\
Ridge Interpolation      & 1.901 & 0.050 \\
\midrule
LAPIS-SHRED              & \textbf{1.065} & \textbf{0.034} \\
\bottomrule
\end{tabular*}
\end{table}

\subsubsection{SINDy-SHRED Comparison (2009 Event)}

SINDy-SHRED~\citep{gao2026sparse} augments the SHRED architecture with a Sparse Identification of Nonlinear Dynamics (SINDy)~\citep{brunton2016discovering} regularization on the latent trajectories, encouraging the learned latent dynamics to conform to parsimonious nonlinear functional forms. This yields an interpretable dynamical system in the latent space---a capability that LAPIS-SHRED does not provide. We evaluate two configurations:

\textbf{SINDy-SHRED Unconstrained.} The discovered SINDy equations are integrated forward from the initial latent state $\mathbf{z}_0$ as a single global rollout, with no intermediate observations. This provides a single initial condition
for forward integration, analogous to the terminal-only frame used by LAPIS-SHRED backward inference, though the unconstrained rollout is susceptible to error accumulation over longer horizons as described in \cref{sec:sindy_method}.

\textbf{SINDy-SHRED Constrained.} The SINDy equations are integrated in overlapping windows of 25 frames, anchored from the ground truth latent trajectory every 5 frames. This requires access to the ground truth latent state at approximately 20\% of all timesteps, providing substantially more temporal information than either LAPIS-SHRED mode.

\Cref{tab:sindy_comparison} presents the results across three scenarios. The comparison reveals a nuanced picture that depends on scenario complexity.

For the short Dry--Dryer--Shallow\,Soil scenario, SINDy-SHRED performs well even in unconstrained mode (RMSE 1.648~m, Corr 0.913), slightly outperforming LAPIS-SHRED (RMSE 1.741~m, Corr 0.907) in RMSE and SSIM. The constrained variant further improves performance (RMSE 1.336~m). This indicates that for compact runout events (i.e., dry, granular flows) with limited temporal extent, the discovered SINDy equations are sufficiently accurate for forward integration without re-initialization.

\begin{figure}[pos=thbp]
\centering
\begin{subfigure}[t]{\columnwidth}
    \centering
    \includegraphics[width=\columnwidth,height=0.35\textheight,keepaspectratio]{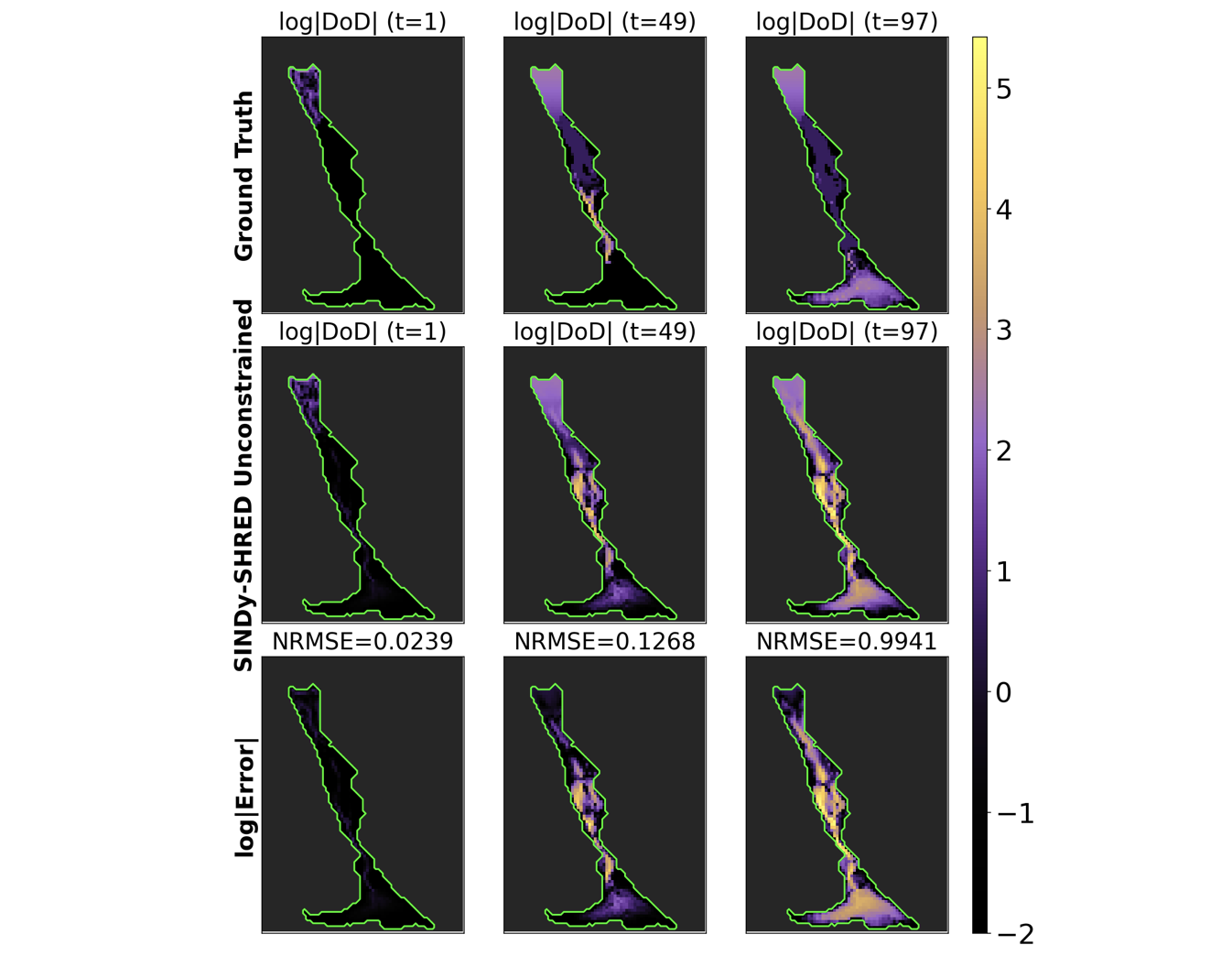}
    \caption{SINDy-SHRED Unconstrained (initial-only observation). RMSE = 12.782~m, Corr = 0.367, SSIM = 0.953.}
    \label{fig:sindy_unconstrained}
\end{subfigure}
\\[6pt]
\begin{subfigure}[t]{\columnwidth}
    \centering
    \includegraphics[width=\columnwidth,height=0.35\textheight,keepaspectratio]{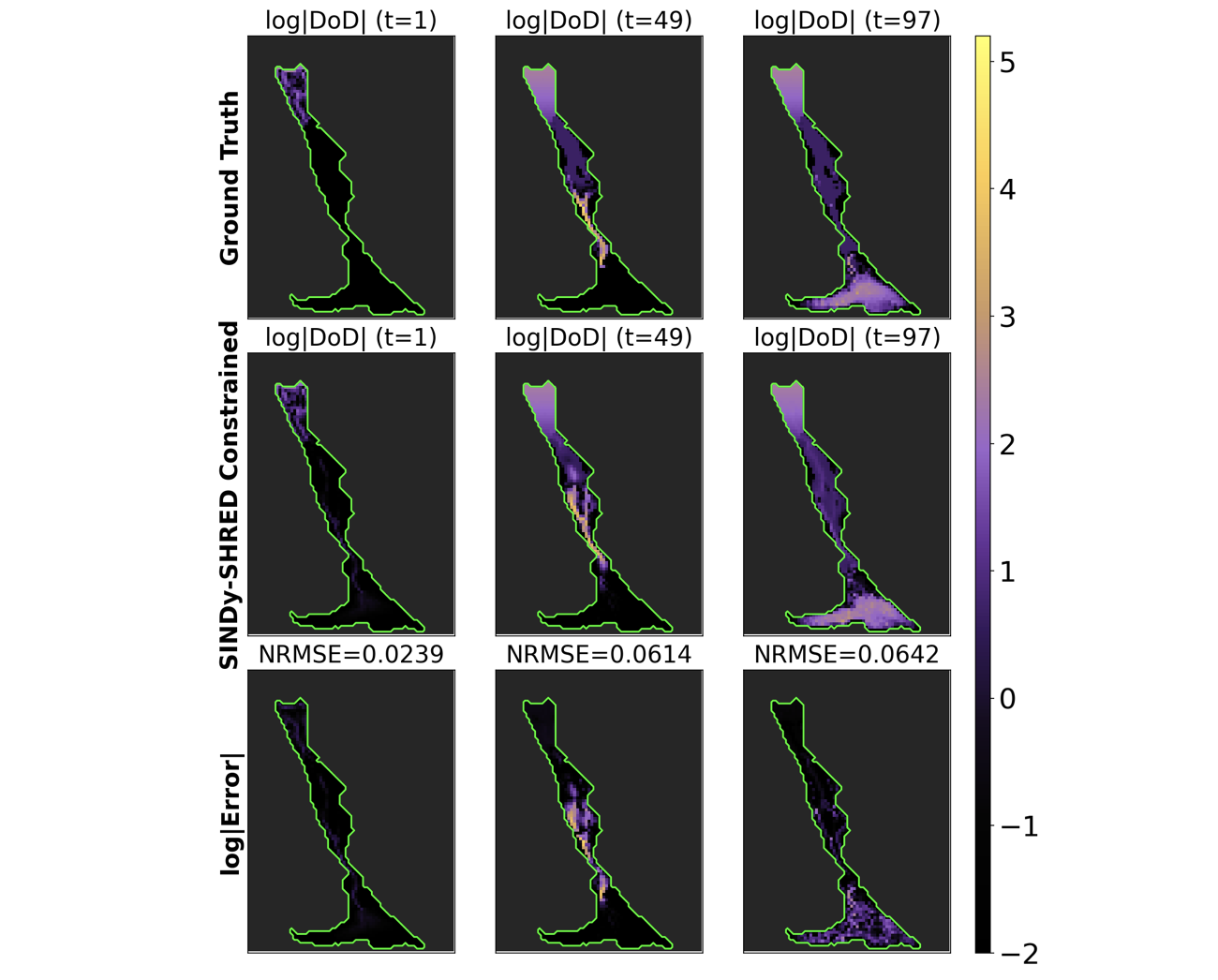}
    \caption{SINDy-SHRED Constrained (anchor observations every 5 frames). RMSE = 3.775~m, Corr = 0.785, SSIM = 0.971.}
    \label{fig:sindy_constrained}
\end{subfigure}
\caption{SINDy-SHRED reconstruction of the 2009 Mount St.\ Helens event under the Wet--Dryer--Deep\,Soil scenario. The unconstrained variant exhibits rapidly growing errors backward from the terminal frame; the constrained variant, with periodic anchor observations, achieves substantially improved reconstruction. Compare with the LAPIS-SHRED reconstruction in \cref{fig:lapis_2009_backward}, which uses only the terminal observation.}
\label{fig:sindy_example}
\end{figure}

For longer and more complex scenarios (i.e., mud-rich slurry-like flows), however, unconstrained SINDy-SHRED degrades substantially. On Moderate--Wetter--Deep\,Soil and Wet--Dryer--Deep\,Soil, RMSE increases to 7.594~m and 12.782~m, respectively---more than 3$\times$ higher than the corresponding LAPIS-SHRED values---and correlation drops below 0.40. The per-frame NRMSE values (\cref{fig:sindy_example}) reveal the mechanism: while frames near the initial condition are reconstructed well (NRMSE $\approx$ 0.02), errors accumulate rapidly along the forward integration, reaching NRMSE of 0.99 by the terminal frame for Wet--Dryer--Deep\,Soil. LAPIS-SHRED, by contrast, maintains 
stable reconstruction quality 
across the full trajectory without requiring equation integration.

SINDy-SHRED Constrained, which anchors from the ground truth latent trajectory every 5 frames, substantially recovers from the unconstrained degradation, achieving RMSE of 3.737--3.775~m on the two longer, complex scenarios. This represents a marked improvement over the unconstrained variant but remains 30--60\% higher in RMSE than LAPIS-SHRED, despite requiring ground truth observations at approximately 20\% of all timesteps---a far richer observation budget than the single terminal frame used by LAPIS-SHRED. In return, SINDy-SHRED provides explicit latent dynamical equations that can be inspected and interpreted, offering mechanistic insight into the latent-space dynamics that LAPIS-SHRED's temporal model does not expose.

This comparison highlights a meaningful tradeoff between observation efficiency and interpretability. 
LAPIS-SHRED achieves strong reconstruction from minimal observations (a single terminal frame or short initial window) but provides no interpretable dynamical equations. 
SINDy-SHRED can discover parsimonious governing equations in the latent space, but requires either short trajectories with limited dynamical complexity (where unconstrained integration remains stable) or substantially denser temporal observations (where periodic anchoring prevents error accumulation) to achieve competitive reconstruction quality. The two approaches are thus complementary: LAPIS-SHRED is suited to operational settings where observations are scarce and rapid reconstruction is the priority, while SINDy-SHRED is suited to physics-oriented analysis where interpretable dynamics are valued and denser observations can be obtained.

\begin{table}[t]
\centering
\caption{SINDy-SHRED comparison on three 2009 Mount St.\ Helens scenarios, ordered from shortest to longest trajectory. SINDy-SHRED Unconstrained integrates the discovered equations forward from a single initial latent state; Constrained anchors from the ground truth latent every 5 frames (using ${\sim}20\%$ of timesteps). LAPIS-SHRED uses only the terminal observation $\mathbf{s}_T$. All RMSE values are computed over the active region.}
\label{tab:sindy_comparison}
\small
\setlength{\tabcolsep}{4pt}
\resizebox{\tblwidth}{!}{%
\begin{tabular}{llccc}
\toprule
\textbf{Scenario} & \textbf{Method} & \textbf{RMSE (m)} & \textbf{Corr} & \textbf{SSIM} \\
\midrule
\multirow{3}{*}{Dry--Dryer--Shallow\,Soil}
  & SINDy-SHRED Unconstrained & 1.648 & 0.913 & 0.989 \\
  & SINDy-SHRED Constrained   & \textbf{1.336} & \textbf{0.936} & \textbf{0.993} \\
  & LAPIS-SHRED (Bwd)         & 1.741 & 0.907 & 0.982 \\
\midrule
\multirow{3}{*}{Moderate--Wetter--Deep\,Soil}
  & SINDy-SHRED Unconstrained & 7.594 & 0.339 & 0.939 \\
  & SINDy-SHRED Constrained   & 3.737 & 0.823 & 0.966 \\
  & LAPIS-SHRED (Bwd)         & \textbf{2.336} & \textbf{0.912} & \textbf{0.982} \\
\midrule
\multirow{3}{*}{Wet--Dryer--Deep\,Soil}
  & SINDy-SHRED Unconstrained & 12.782 & 0.367 & 0.953 \\
  & SINDy-SHRED Constrained   & 3.775 & 0.785 & 0.971 \\
  & LAPIS-SHRED (Bwd)         & \textbf{2.878} & \textbf{0.884} & \textbf{0.975} \\
\bottomrule
\end{tabular}%
}
\end{table}

\subsection{Sensitivity and Ablation Studies}
\label{sec:sensitivity}

We investigate the sensitivity of LAPIS-SHRED with respect to three factors: DEM spatial resolution, sensor placement strategy, and the breadth of the simulation prior.
Sensitivity to DEM resolution is evaluated on the 2009 event, where
topographic controls on the lateral spread of the runout path
and multiple scenario regimes make resolution effects on flow routing most apparent; sensitivities to sensor placement and parameter perturbation are evaluated on the 2022 event, a distinct failure configuration that provides complementary validation beyond the 2009 calibration scenarios.

\subsubsection{DEM Resolution}
\label{sec:dem_resolution}

The spatial resolution of the input DEM governs the fidelity with which terrain features are represented in the MWR simulation. We evaluate LAPIS-SHRED backward inference on the Wet--Dryer--Deep\,Soil scenario of the 2009 event at four DEM resolutions: 5~m, 10~m (native), 20~m, and 30~m. The native 10~m DEM is resampled to the target resolutions using bilinear interpolation (i.e., terrain features in the 5m DEM are identical to the 10 m DEM but represented by 4 times as many grid cells), and all MWR simulations and LAPIS-SHRED training are conducted independently at each resolution. The sensor count ($p = 32$) and all architectural hyperparameters are held constant.

\begin{figure*}[t]
\centering
\includegraphics[width=\textwidth]{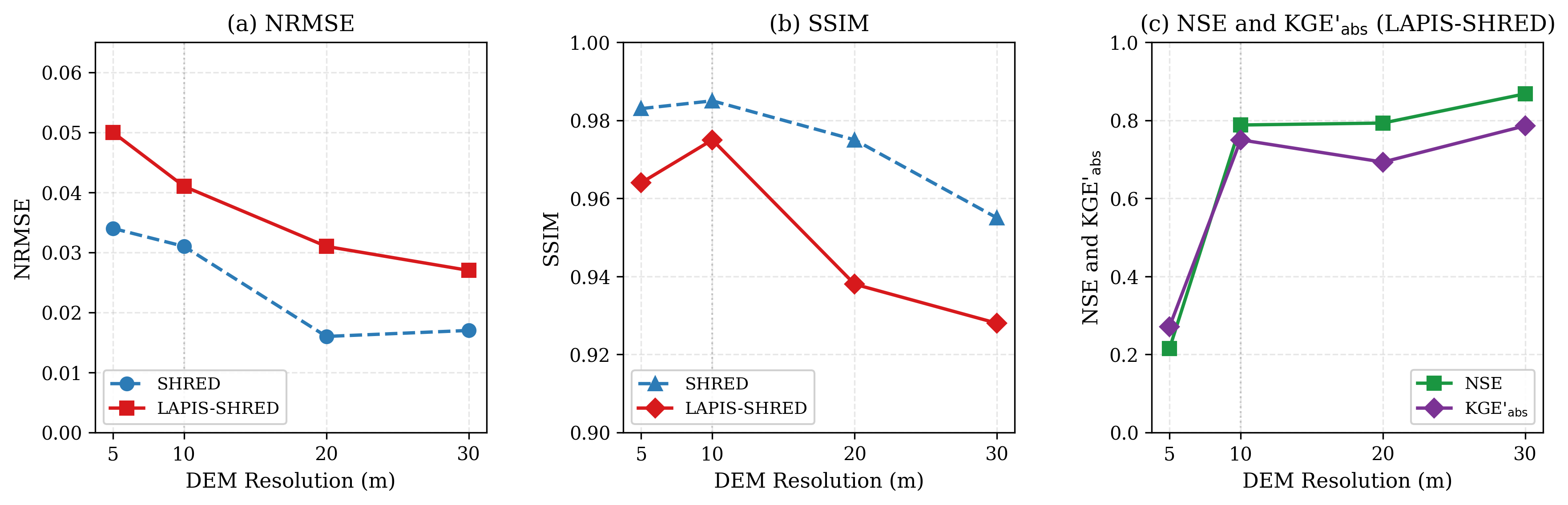}
\caption{Effect of DEM resolution on LAPIS-SHRED backward inference for the Wet--Dryer--Deep\,Soil scenario (2009 event). (a)~NRMSE and (b)~SSIM for both SHRED (upper bound, dashed lines) and LAPIS-SHRED (solid lines). (c)~NSE and KGE$'_{\text{abs}}$ for LAPIS-SHRED. The native 10~m resolution is indicated by the vertical dotted line.}
\label{fig:dem_resolution_metrics}
\end{figure*}

\begin{figure*}[pos=thbp]
\centering
\begin{subfigure}[t]{\columnwidth}
    \centering
    \includegraphics[width=\columnwidth]{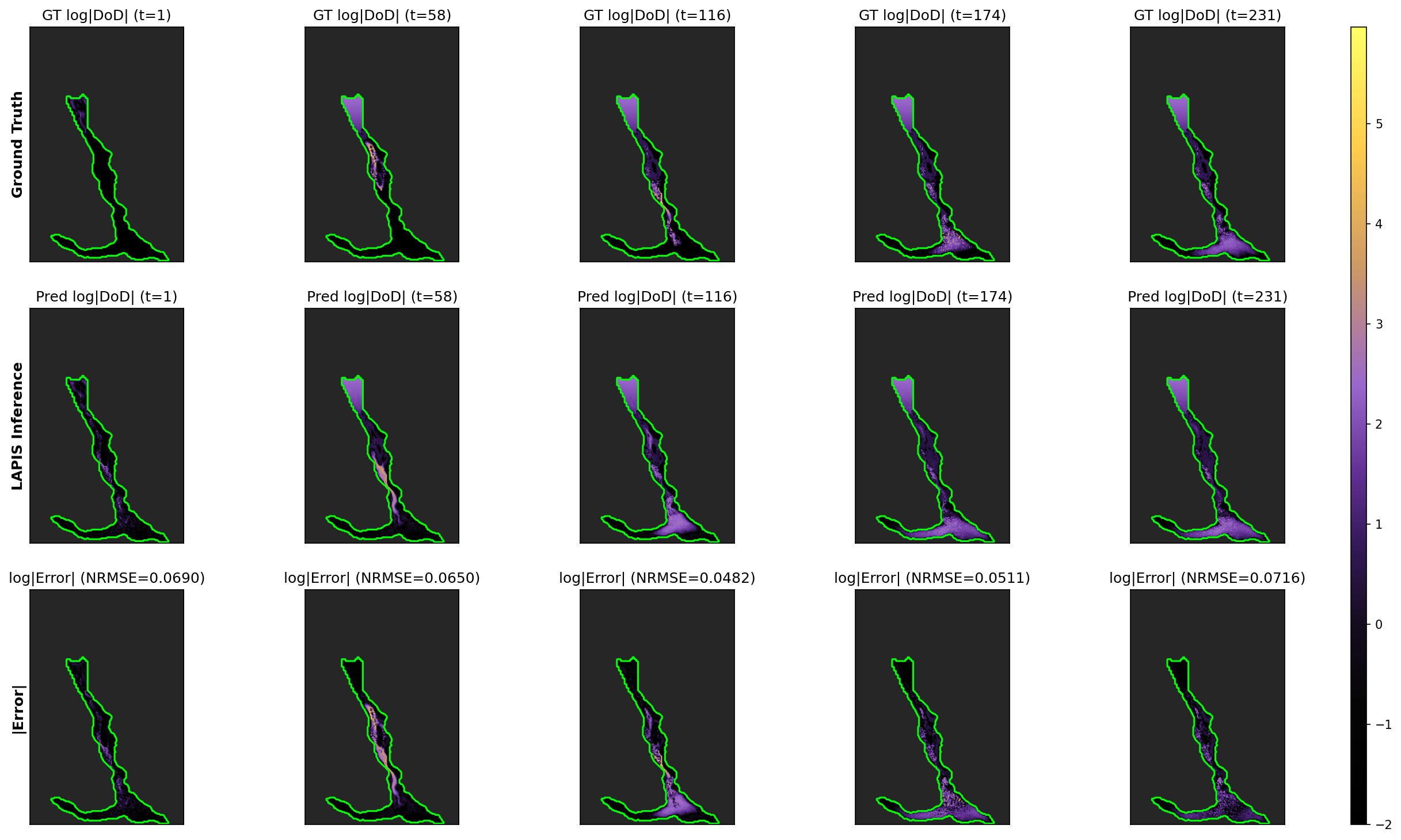}
    \caption{5~m resolution.}
\end{subfigure}
\hfill
\begin{subfigure}[t]{\columnwidth}
    \centering
    \includegraphics[width=\columnwidth]{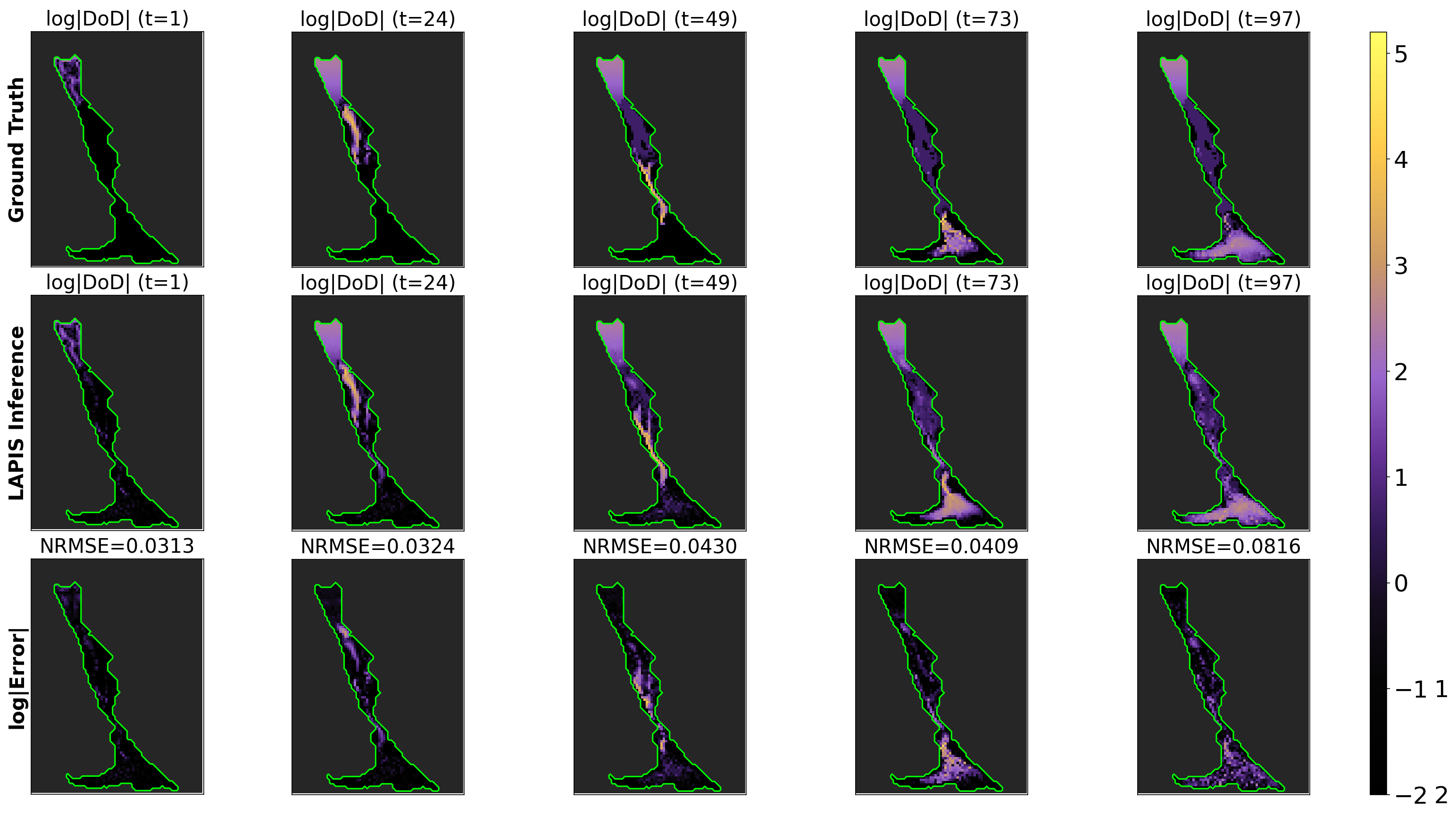}
    \caption{10~m resolution (native).}
\end{subfigure}
\\[6pt]
\begin{subfigure}[t]{\columnwidth}
    \centering
    \includegraphics[width=\columnwidth]{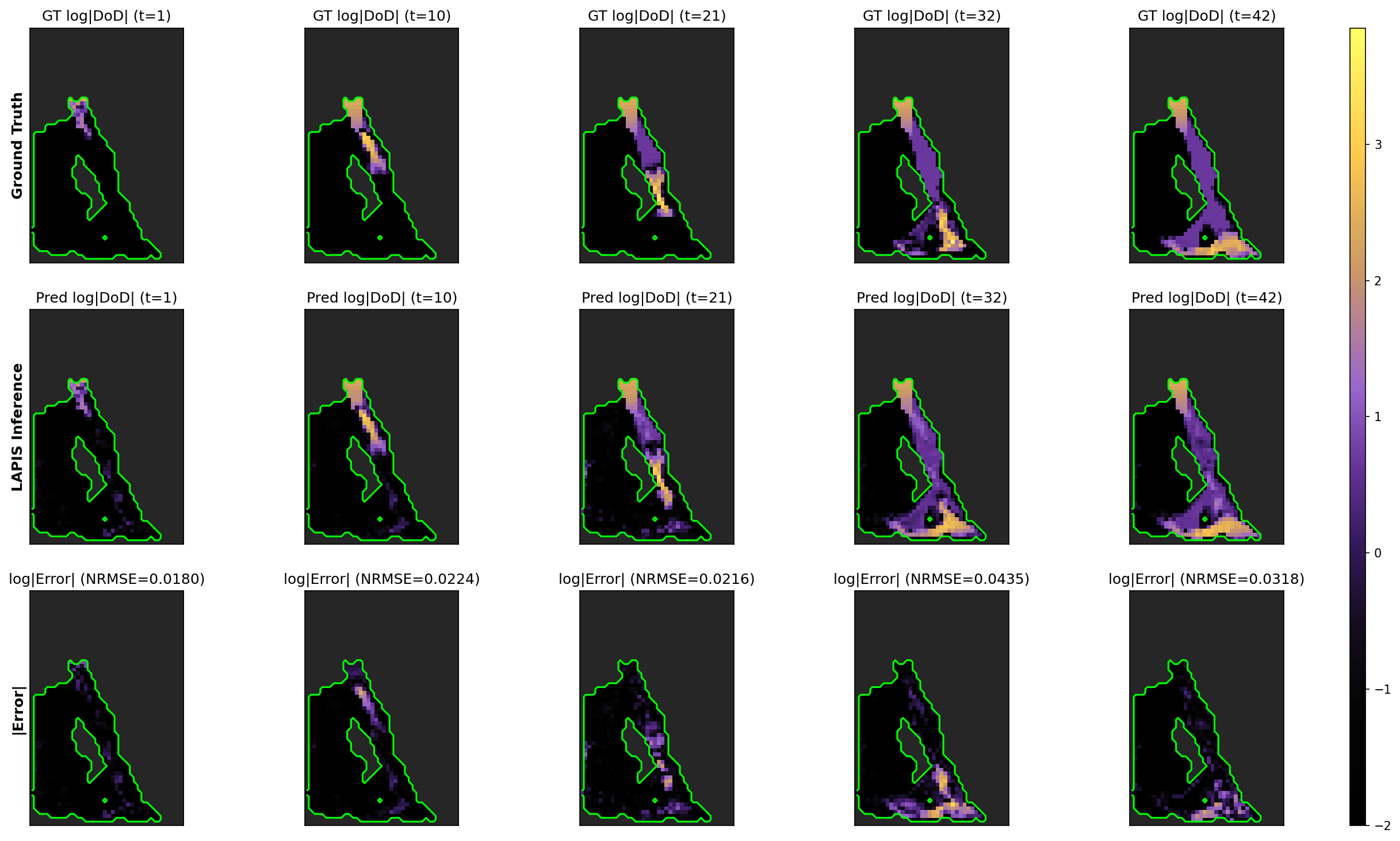}
    \caption{20~m resolution.}
\end{subfigure}
\hfill
\begin{subfigure}[t]{\columnwidth}
    \centering
    \includegraphics[width=\columnwidth]{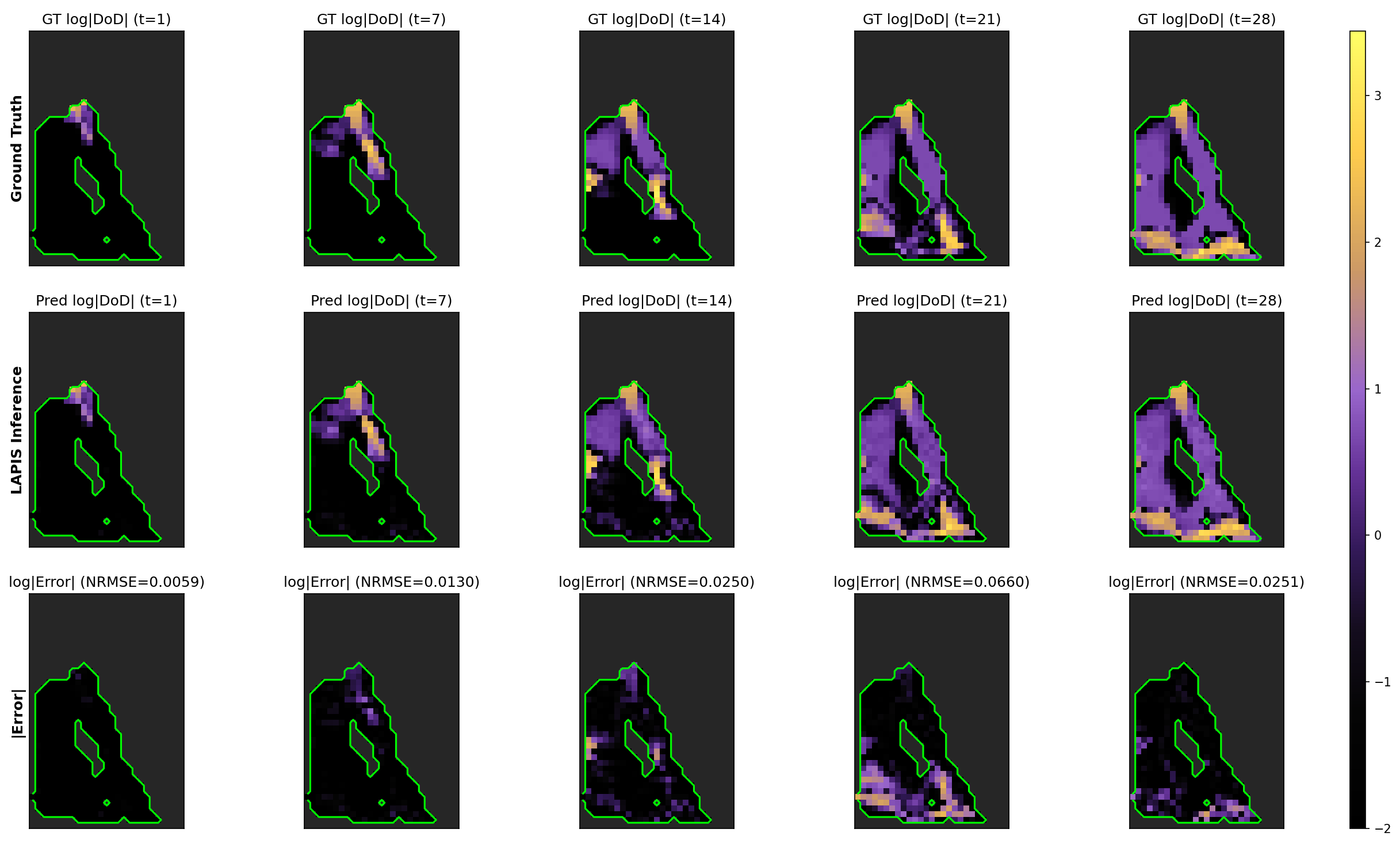}
    \caption{30~m resolution.}
\end{subfigure}
\caption{LAPIS-SHRED backward reconstruction of the Wet--Dryer--Deep\,Soil scenario at four DEM resolutions. Each panel shows ground truth DoD (top row), LAPIS-SHRED reconstruction (middle row), and error (bottom row) at selected timesteps. Coarser resolutions yield lower NRMSE but lose terrain features that control flow routing, producing qualitatively different runout patterns. Log-scale visualization; green contours outline the active region.}
\label{fig:dem_resolution}
\end{figure*}

\Cref{fig:dem_resolution_metrics} summarizes the resolution-sensitivity results. Coarser grids yield lower NRMSE and higher NSE, with the 30~m grid attaining the best numerical values (NRMSE~=~0.027, NSE~=~0.868) and the 5~m grid the weakest LAPIS-SHRED scores (NRMSE~=~0.050, NSE~=~0.215). This apparent improvement at coarser resolution, however, should be interpreted cautiously. As shown in \Cref{fig:dem_resolution}, DEM smoothing at 20~m and 30~m removes fine-scale terrain barriers that are necessary to
constrain lateral spreading in
parameterizations with larger runout extents, producing runout patterns with spatial extents that differ substantially from those at the native resolution. The decline in SSIM at coarser grids (\Cref{fig:dem_resolution_metrics}a) is consistent with this loss of spatial structure. Thus, the lower NRMSE primarily reflects the reduced complexity of the smoothed DoD field rather than improved physical fidelity.

The 5~m case represents the opposite limit: finer resolution increases both spatial dimensionality (${\sim}4{\times}$ more grid cells than the 10~m case) and temporal complexity, with MWR trajectories extending to 231 frames compared with 97 frames at 10~m. Reconstruction accuracy consequently decreases for both LAPIS-SHRED and the SHRED upper bound: NRMSE increases from 0.041 to 0.050 for LAPIS-SHRED and from 0.031 to 0.034 for SHRED. This shared degradation indicates that the higher-resolution domain is intrinsically more difficult to reconstruct, rather than the performance loss arising solely from the backward inference mechanism.

We note that these resolutions are derived via interpolation and thus do not constitute a definitive assessment of optimal resolution. Nevertheless, the native 10~m resolution provides a reasonable balance: it preserves terrain features relevant to flow routing while remaining tractable for the LAPIS-SHRED architecture.

\subsubsection{Sensor Placement}
\label{sec:sensor_placement}

\begin{figure}[pos=thbp]
\centering
\begin{subfigure}[t]{0.48\columnwidth}
    \centering
    \includegraphics[width=\columnwidth]{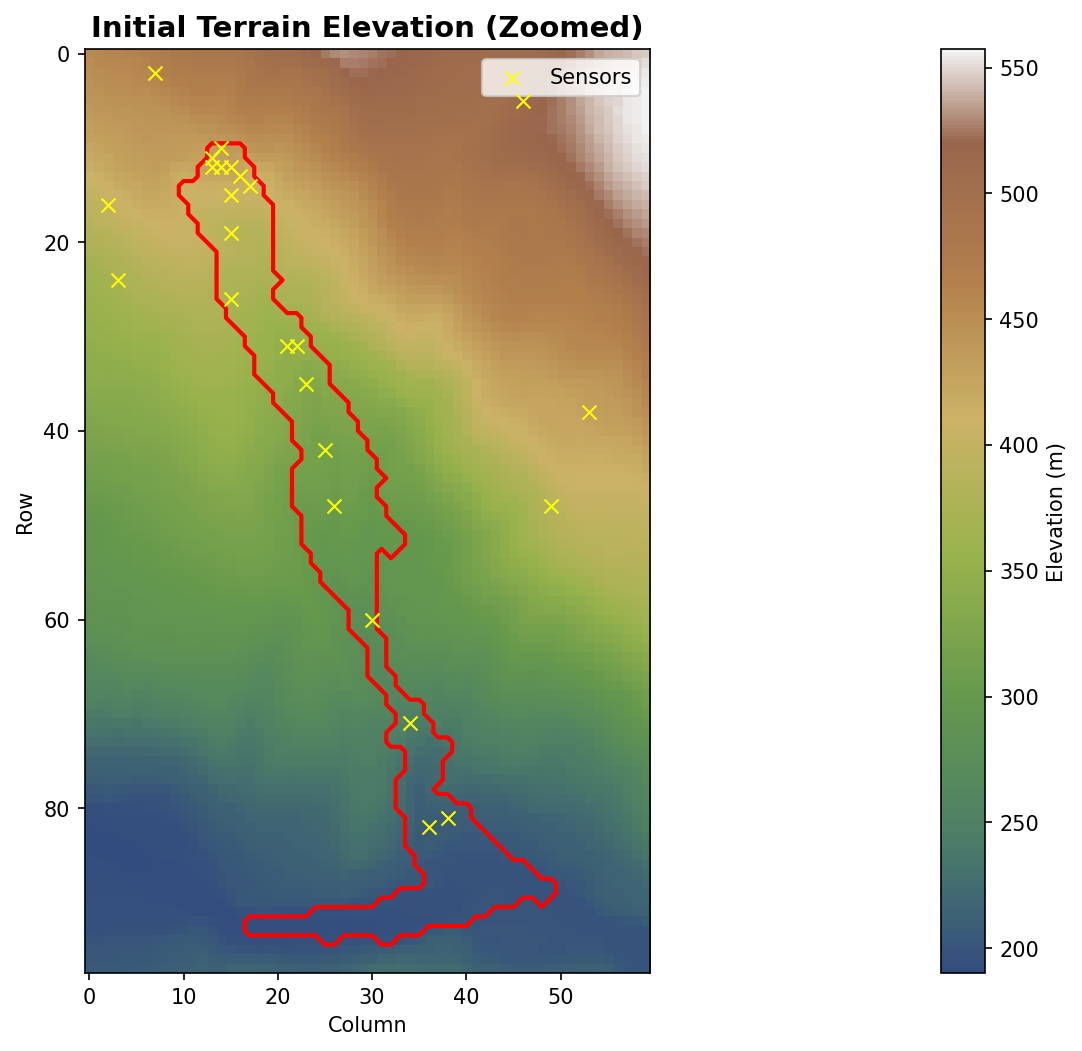}
    \caption{Default (DoD-weighted).}
\end{subfigure}
\hfill
\begin{subfigure}[t]{0.48\columnwidth}
    \centering
    \includegraphics[width=\columnwidth]{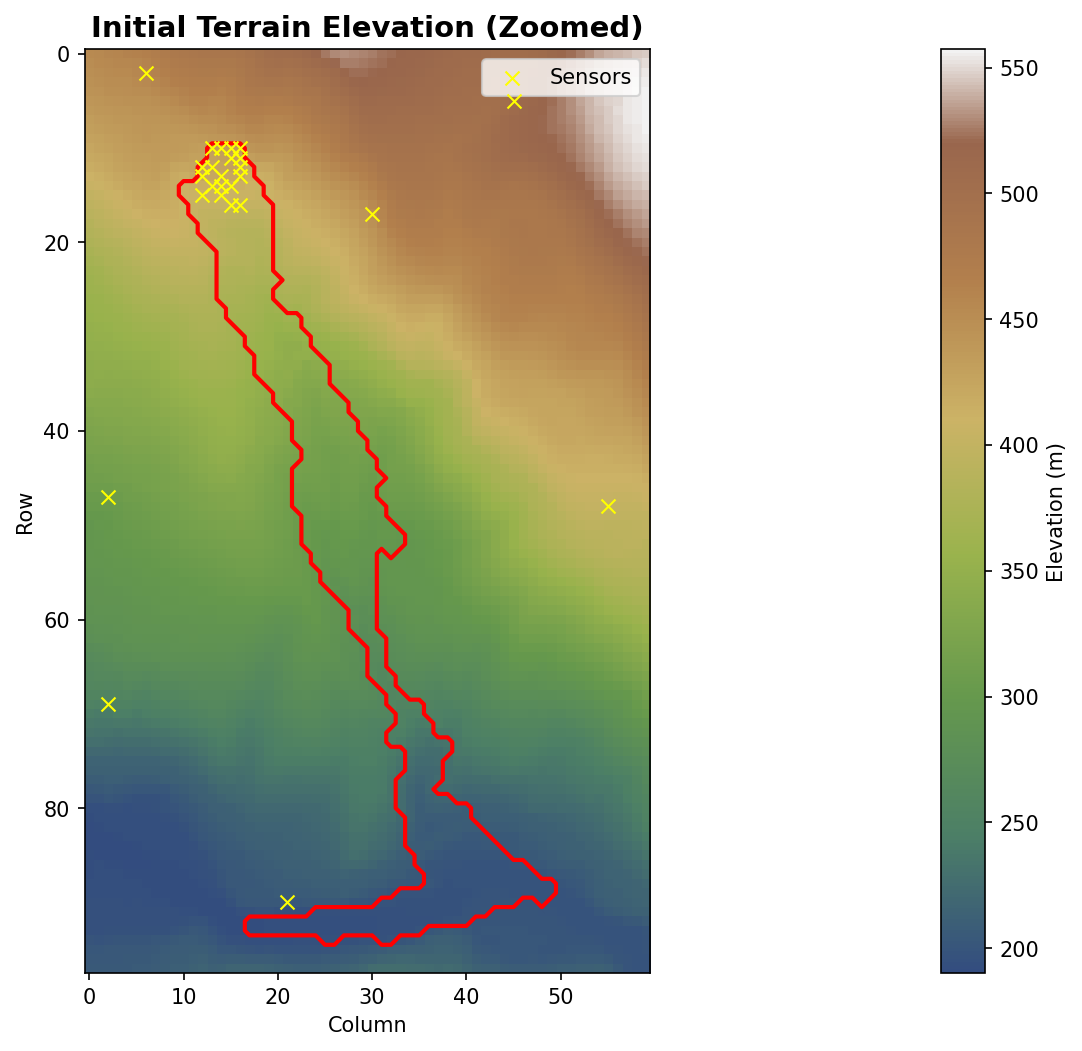}
    \caption{Top-clustering.}
\end{subfigure}
\caption{Sensor placement configurations overlaid on the initial terrain elevation (simulated 2022 event, zoomed view). Yellow markers indicate sensor locations; the red contour outlines the landslide runout boundary; background shows initial terrain elevation.}
\label{fig:sensor_placement}
\end{figure}

We evaluate two sensor placement strategies on the 2022 event to assess the sensitivity of reconstruction quality to observation locations: (i)~the \emph{default} DoD-weighted configuration, where sensors within the active region are distributed proportionally to simulation-inferred DoD magnitudes along the runout path; and (ii)~a \emph{top-clustering} configuration, where all 18 active-region sensors are concentrated at the landslide initiation zone, simulating pre-event deployment constraints. The remaining 14 sensors are distributed across quiescent regions in both configurations. \Cref{fig:sensor_placement} illustrates both layouts.

\Cref{tab:sensor_placement} presents the backward inference results. The default placement achieves RMSE of 1.065~m, while top-clustering yields 1.264~m---a 19\% increase. For comparison, the SHRED upper bound degrades more steeply under top-clustering (RMSE increasing from 0.867~m to 1.132~m, a 31\% increase), suggesting that the LAPIS-SHRED backward inference mechanism provides some robustness to sensor placement by leveraging learned temporal dynamics to compensate for reduced spatial coverage.

\begin{table}[t]
\centering
\caption{Effect of sensor placement on LAPIS-SHRED backward inference (2022 event, broad perturbation regime). Default: DoD-weighted along runout path; Top-clustering: all active-region sensors at the initiation zone.}
\label{tab:sensor_placement}
\small
\resizebox{\tblwidth}{!}{%
\begin{tabular}{lcccc}
\toprule
& \multicolumn{2}{c}{\textbf{SHRED (upper bound)}} & \multicolumn{2}{c}{\textbf{LAPIS-SHRED}} \\
\cmidrule(lr){2-3} \cmidrule(lr){4-5}
\textbf{Placement} & \textbf{RMSE (m)} & \textbf{SSIM} & \textbf{RMSE (m)} & \textbf{SSIM} \\
\midrule
Default (DoD-weighted)  & 0.867 & 0.981 & 1.065 & 0.979 \\
Top-clustering          & 1.132 & 0.966 & 1.264 & 0.948 \\
\bottomrule
\end{tabular}%
}
\end{table}

\subsubsection{Sensitivity to Prior Parameter Precision}
\label{sec:perturbation_range}

LAPIS-SHRED is trained on simulation ensembles intended to span plausible parameter variability. To assess sensitivity to the precision of this prior, we compare two perturbation regimes for the 2022 event: a broad regime ($\pm 100\%$, $K = 15$ simulations), representing limited prior constraint, and a narrow regime ($\pm 10$--$15\%$, $K = 6$ simulations), representing more tightly calibrated parameter estimates.

\Cref{tab:perturbation_range} summarizes the comparison for the default sensor configuration. Under the narrow regime, LAPIS-SHRED achieves similar RMSE under the narrow and broad regimes (1.052~m and 1.065~m, respectively), indicating that reconstruction accuracy is largely insensitive to the perturbation range over the configurations considered. In particular, expanding the simulation ensemble from a tightly constrained parameter neighborhood to a substantially broader range does not materially degrade performance. While the ensemble sizes differ between regimes, the comparable accuracy suggests that coverage of the parameter space, rather than precise localization around the ground truth, is sufficient for effective reconstruction in this setting. In practice, the larger ensemble size associated with broader perturbations is not prohibitive, as simulation data can be generated offline for pretraining prior to deployment. 

This behavior is consistent with the modular structure of LAPIS-SHRED: the spatial reconstruction module learns a latent representation from simulations sharing the same terrain geometry, while the temporal model learns how trajectories evolve within this representation. The perturbed parameters affect erosion--deposition behavior, flow arrest, entrainment, and available mobilizable material, but do not alter the underlying DEM. As a result, the ensemble retains a coherent spatial organization across parameter configurations, enabling the SHRED representation and backward temporal model to generalize across the examined range without requiring the training ensemble to be tightly localized around the held-out event.

\begin{table}[t]
\centering
\caption{Comparison of LAPIS-SHRED performance across perturbation regimes (2022 event, default sensor configuration). The narrow regime uses $K = 6$ simulations with $\pm 10$--$15\%$ perturbations; the broad regime uses $K = 15$ simulations with $\pm 100\%$ perturbations.}
\label{tab:perturbation_range}
\small
\resizebox{\tblwidth}{!}{%
\begin{tabular}{lccccc}
\toprule
& & \multicolumn{2}{c}{\textbf{SHRED (upper bound)}} & \multicolumn{2}{c}{\textbf{LAPIS-SHRED}} \\
\cmidrule(lr){3-4} \cmidrule(lr){5-6}
\textbf{Regime} & $K$ & \textbf{RMSE (m)} & \textbf{SSIM} & \textbf{RMSE (m)} & \textbf{SSIM} \\
\midrule
Narrow ($\pm 10$--$15\%$)  & 6  & 0.758 & 0.981 & 1.052 & 0.961 \\
Broad ($\pm 100\%$)        & 15 & 0.867 & 0.981 & 1.065 & 0.979 \\
\bottomrule
\end{tabular}%
}
\end{table}

\subsection{Validation Against Field-Observed Deposits}
\label{sec:observed_validation}

\begin{figure}[pos=thbp]
\centering
\includegraphics[width=\columnwidth]{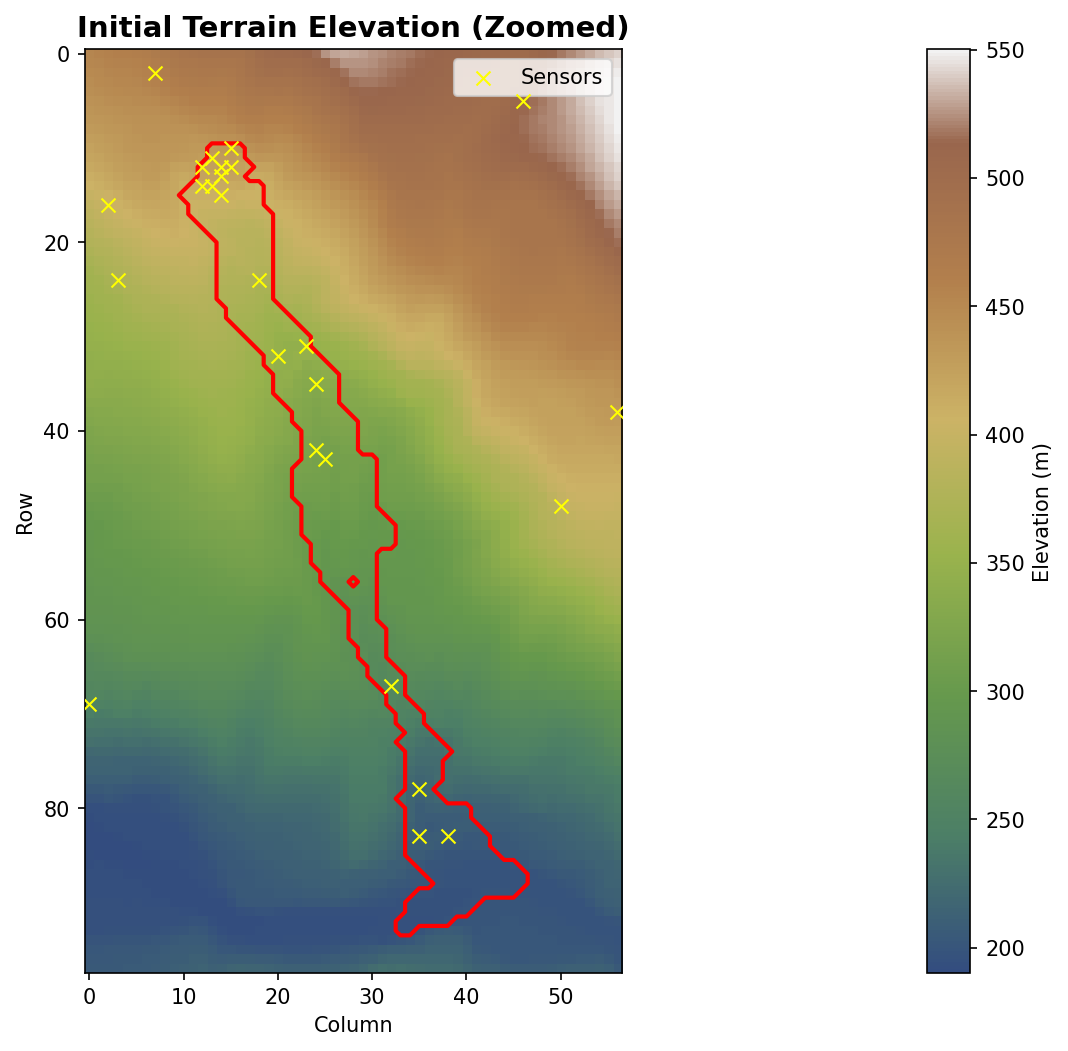}
\caption{Sensor placement for the field-validation experiment. Yellow markers indicate sensor locations, placed using the observed LiDAR terminal DoD following the active/quiescent allocation described in \cref{sec:sensor_placement}. The red contour outlines the landslide runout boundary; background shows initial terrain elevation.}
\label{fig:observed_sensors}
\end{figure}

\begin{figure*}[pos=thbp]
\centering
\includegraphics[width=\textwidth]{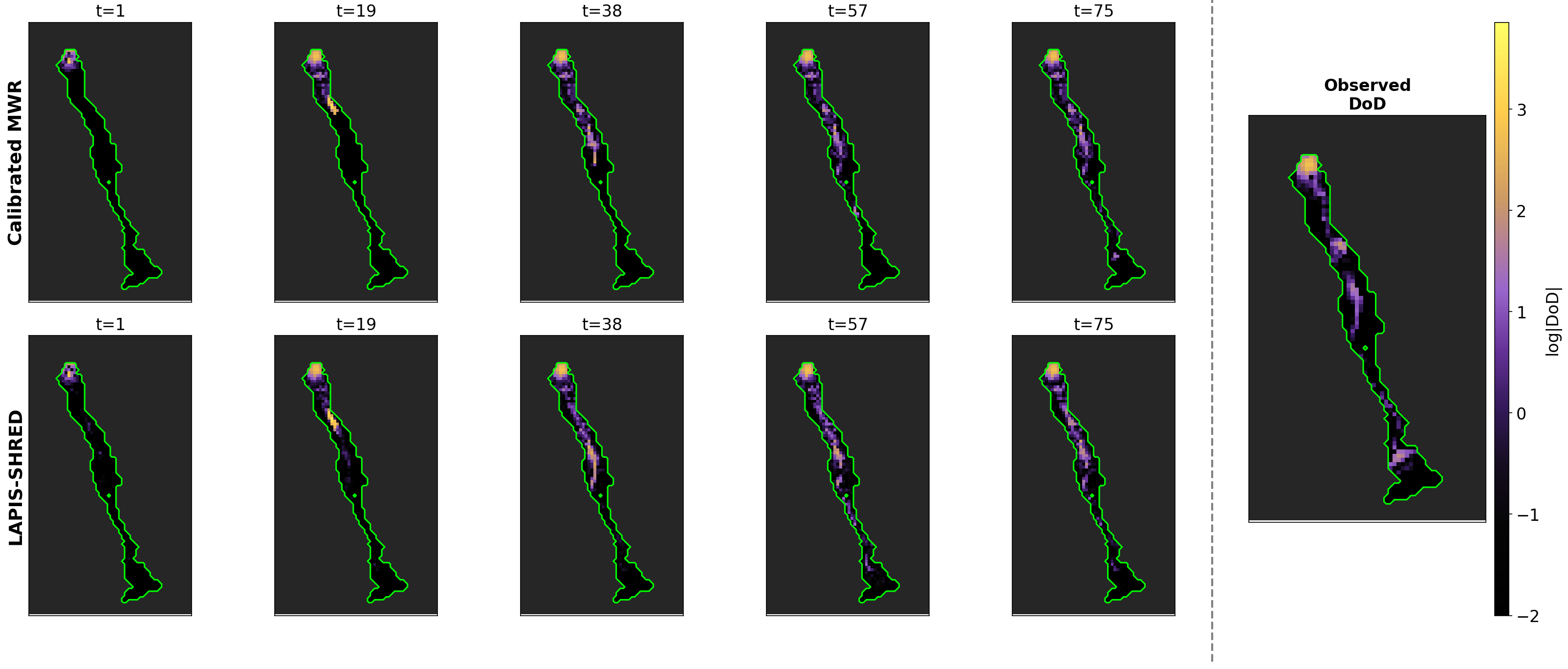}
\caption{Validation against LiDAR-observed actual deposits (2022 event, $K = 25$ ensemble). Row~1: Calibrated MWR-simulated trajectory at selected timesteps; Row~2: LAPIS-SHRED backward reconstruction from terminal observation. The rightmost column shows the observed actual terminal DoD. Log-scale visualization; green contours outline the active region.}
\label{fig:observed_validation}
\end{figure*}

The preceding evaluations use simulated ground truth, for which the complete spatiotemporal trajectory is available for held-out quantitative assessment. In contrast, actual field observations of rapid runout events rarely capture the transient runout process; typically, only the post-event deposit could be documented. We therefore examine whether LAPIS-SHRED, trained entirely on simulated ensembles, can efficiently recover a plausible trajectory for real runout deposits observed in the field. To this end, we train LAPIS-SHRED on an ensemble of $K = 25$ MWR simulations of the January 2022 retrogressive enlargement using broad parameter ranges representative of plausible event conditions ($S_c \in [0.095,\, 0.225]$, $q_c \in [0.42,\, 1.14]$, $k \in [0.012,\, 0.058]$), and evaluate the reconstructed terminal deposit against the LiDAR-derived terminal DoD from that event. At inference, LAPIS-SHRED receives only sensor observations sampled from the post-event LiDAR DoD field and reconstructs the full spatiotemporal trajectory backward.
\Cref{fig:observed_sensors} shows the sensor placement on the observed deposit footprint.

For comparison, an event-specific MWR parameterization was independently obtained through Bayesian MCMC calibration to the 2022 observations: $S_c = 0.155$, $q_c = 0.76$, $k = 0.034$ (${\sim}2{,}000$ forward evaluations; \citealp{keck2024landslide}). This calibrated MWR simulation provides a complementary physics-based reference trajectory and is not used to define or train the LAPIS-SHRED training ensemble. Notably, the ensemble ranges are broad relative to the calibrated values, extending approximately 39--71\% away from the calibrated values across the three parameters.

\begin{table}[t]
\centering
\caption{Terminal-frame Pearson correlation ($Corr$) and active-region RMSE against the 2022 event's actual observed deposits ($K = 25$ ensemble spanning $\pm 39$--$71\%$ of calibrated parameters).}
\label{tab:observed_validation}
\small
\begin{tabular*}{\tblwidth}{@{\extracolsep{\fill}}lcc@{}}
\toprule
\textbf{Method} & $Corr$ & RMSE (m) \\
\midrule
MWR (calibrated $\boldsymbol{\theta}^*$) & 0.853 & 1.630 \\
SHRED (upper bound, $\{\mathbf{s}_1, \ldots, \mathbf{s}_T\}$) & 0.853 & 1.269 \\
LAPIS-SHRED ($\mathbf{s}_T$ only)                 & 0.843 & 1.331 \\
\bottomrule
\end{tabular*}
\end{table}

\Cref{fig:observed_validation} illustrates the calibrated MWR-simulated trajectory and the LAPIS-SHRED-reconstructed trajectory together with the observed terminal deposit, enabling comparison of their terminal states against the LiDAR-derived DoD.
\Cref{tab:observed_validation} compares the terminal deposits predicted by the calibrated MWR simulator, the SHRED upper bound (full trajectory sensor observation), and LAPIS-SHRED (terminal frame sensor observation only) against the observed deposited DoD. The calibrated MWR achieves Corr 0.853; SHRED assuming full temporal sequence also achieves Corr 0.853, and LAPIS-SHRED achieves Corr 0.843 from sparse sensor observations of the terminal frame only. Active-region RMSE values are 1.630, 1.269, and 1.331~m, respectively. Overall, LAPIS-SHRED achieves performance comparable to the calibrated MWR simulation despite using only sparse terminal-frame observations at inference.

\begin{table}[t]
\centering
\caption{Computational cost comparison for the 2022 event on a $122 \times 80$ cell grid. MWR calibration evaluations are sequential (MCMC); LAPIS-SHRED simulations are independent and trivially parallelizable.}
\label{tab:computational_cost}
\small
\begin{tabular*}{\tblwidth}{@{\extracolsep{\fill}}lcc@{}}
\toprule
 & \textbf{MWR Calibration} & \textbf{LAPIS-SHRED} \\
\midrule
Forward evaluations         & ${\sim}2{,}000$ (sequential)  & $25$ (independent) \\
Per-evaluation time         & ${\sim}0.6$~s                 & ${\sim}0.6$~s \\
Simulation wall time        & ${\sim}20$~min                & ${\sim}15$~s \\
Model training              & ---                           & ${\sim}3.5$~min \\
\midrule
\textbf{Total wall time}    & $\boldsymbol{{\sim}20}$~\textbf{min} & $\boldsymbol{<4}$~\textbf{min} \\
\midrule
\multicolumn{3}{c}{\textit{Hardware}: Apple M4, 24~GB RAM} \\
\bottomrule
\end{tabular*}
\end{table}

The MWR calibration required ${\sim}2{,}000$ sequential MWR forward evaluations against an observed runout (\cref{sec:data_generation}). On the study grid ($122 \times 80$ cells) and the hardware used in this study (Apple M4, 24~GB RAM), each forward evaluation completes in ${\sim}0.6$~s, placing the sequential calibration cost at ${\sim}20$~min; the MCMC chain is sequential, as each proposal depends on the preceding evaluation. The LAPIS-SHRED pipeline requires $K = 25$ independent forward simulations (could be parallelized) only and ${\sim}3.5$~min of model training, costing less than 4~min of wall-clock time to produce a reconstruction from broad parameter bounds and a single sparse post-event observation (\cref{tab:computational_cost}). 
This cost difference would become more pronounced for larger or higher-resolution domains, particularly because the LAPIS-SHRED training ensemble uses substantially fewer simulations that can be executed in parallel, whereas the MCMC calibration procedure used here proceeds sequentially.

Internal consistency with the simulated ground truth over the full trajectory yields NRMSE of 0.029 and SSIM of 0.980 for LAPIS-SHRED (SHRED: 0.022, 0.988), comparable to \cref{tab:lapis_2021}. These results indicate that the reconstruction is not limited to reproducing the observed terminal state, but also remains consistent with the spatiotemporal evolution by the calibrated physics-based MWR model used as a reference in this study. Evaluation against field observations is limited to the terminal frame because no intermediate ground observations of the 2022 event are available.

\section{Discussion}
\label{sec:discussion}

The experimental results establish that LAPIS-SHRED can reconstruct landslide runout dynamics from severely constrained observations across a range of scenario conditions. In this section, we discuss implications for operational geohazard assessment, practical deployment considerations arising from the sensitivity studies, and limitations of the current approach.

\subsection{Operational Implications for Landslide Assessment}
\label{sec:operational}

The results of \cref{sec:results} demonstrate that LAPIS-SHRED supports two complementary modes of landslide runout reconstruction, each aligned with a distinct operational context.

\textbf{Post-event reconstruction.} Backward inference from a single terminal observation corresponds to settings in which post-event information is available---for example, from field surveys, aerial imagery, satellite remote sensing, or LiDAR/DEM differencing---while the transient flow dynamics were not directly observed. As shown in \cref{sec:results_core}, LAPIS-SHRED reconstructs the full spatiotemporal evolution from terminal-only observations, attaining NRMSE below 5.8\% across
scenarios spanning dry to wet climates and shallow to deep soil conditions. This capability could support forensic analysis of past events, inform calibration of physics-based runout models, and provide temporal context for interpreting observed deposit morphologies.
The field validation in \cref{sec:observed_validation} provides initial evidence that this capability extends to operational settings with real post-event observations.

\textbf{Early-warning and scenario exploration.} Forward inference from a short initial observation window corresponds to settings in which early deformation or failure-related signals are available from monitoring systems~\citep{intrieri2013brief}, such as seismic sensors, tiltmeters/inclinometers, GNSS, or InSAR~\citep{scaioni2014remote}. The comparable performance of forward and backward modes observed in \cref{sec:results_core} suggests that both are viable for operational use, though the relative informativeness of initial versus terminal observations depends on the specific runout regime.

Recent cascading high-mountain disasters underscore the potential value of such forward-inference capabilities. On 26 August 2026 near the China--Nepal border, 
a complex debris flow initiated by a bedrock landslide and ice avalanche---rather than by rainfall or sudden snowmelt---propagated tens of kilometers through the downstream river corridor, 
while retrospective satellite analysis revealed substantial pre-failure deformation and an accelerating phase preceding the collapse \citep{wang2026precursor}. This event illustrates a broader operational challenge: detecting precursor deformation or the onset of failure does not, by itself, provide a prompt and accurate forecast of the subsequent runout and downstream impacts. Given site-specific simulations generated from physics-based models under plausible parameter regimes, a trained forward LAPIS-SHRED model could rapidly map the earliest available measurements of an unfolding event to projected impact trajectories, thereby supporting time-sensitive hazard assessment and emergency response.
A remaining challenge for deploying such frameworks at ungauged or poorly characterized sites is the specification of plausible parameter regimes for generating the underlying simulation ensembles. Emerging capabilities of large-language-model agents in translating qualitative geological and geomorphological descriptions into structured geotechnical parameter estimates \citep{wu2025perspectives} suggest one possibility for automating this step, though systematic benchmarking against expert judgment and field measurements will be required.

Beyond these operational interpretations, both inference modes can be used for scenario-based analysis. Backward inference enables reconstruction of plausible antecedent trajectories conditioned on a specified terminal deposit, while forward inference allows projection of runout evolution from a prescribed initial failure configuration. These capabilities support rapid exploration of alternative scenarios under the learned simulation prior. Such analyses could complement physics-based simulations, suggesting a potential role for LAPIS-SHRED as a flexible tool for exploring landslide dynamics under data-constrained conditions.

\subsection{Practical Considerations}
\label{sec:practical}

The sensitivity analyses in \cref{sec:sensitivity} highlight several considerations for applying LAPIS-SHRED to landslide reconstruction.

\textbf{DEM resolution.} The resolution of the input DEM plays a critical role in balancing numerical performance and physical fidelity. Coarser grids tend to yield better metrics due to reduced spatial complexity, but may suppress terrain features that govern flow routing. This could lead to reconstructions that are metrically accurate yet physically inconsistent with the underlying landscape. For the Mt.\ St.\ Helens case studied, the native 10~m DEM provides a suitable working resolution, preserving key topographic controls while remaining computationally tractable. More generally, DEM resolution should be selected in accordance with the characteristic scale of terrain features that influence runout behavior.

\textbf{Sensor placement.} The finding that LAPIS-SHRED degrades less than the SHRED upper bound under top-clustering (19\% vs.\ 31\% RMSE increase) suggests that the temporal model partially compensates for spatial information loss by leveraging learned dynamical structure. This has implications for monitoring network design in landslide-prone areas: in practice, pre-event sensor deployments are often concentrated near the headscarp or anticipated initiation zone, where hazard is most readily identified, while coverage along the eventual runout path is typically limited. The results indicate that such constrained deployments, while suboptimal, remain viable for LAPIS-SHRED reconstruction.

\textbf{Simulation ensemble design.} The comparison between broad and narrow parameter perturbation regimes indicates that LAPIS-SHRED reconstruction accuracy is relatively insensitive to the precision of prior parameter estimates in the tested setting. In many operational settings, geotechnical parameters are known only to order-of-magnitude precision. The result indicates that LAPIS-SHRED can tolerate this level of uncertainty, provided the true parameters fall within the ensemble's convex hull.

\textbf{Computational cost.} The lightweight, modular architecture of LAPIS-SHRED enables a deployment model in which computationally expensive steps---including ensemble simulation and SHRED training across flow regimes spanning granular to slurry-like behavior---can be performed offline during hazard assessment planning, while inference from new observations requires only a single evaluation of the trained model, with minimal computational overhead on conventional hardware. This separation between offline preparation and real-time inference is well suited to operational geohazard workflows, where pre-computed scenario libraries are maintained and rapid assessment is needed when an event occurs or is anticipated. 
For example, for the 2022 field validation (\cref{sec:observed_validation}), the complete LAPIS-SHRED pipeline ran in under 4~min on an Apple M4 processor, compared with ${\sim}20$~min for the sequential MWR calibration procedure on the same grid. This computational cost difference becomes more pronounced on larger or higher-resolution domains, where individual physics-based model evaluations are more computationally demanding.

\subsection{Limitations}
\label{sec:limitations}

The present study employs the MWR cellular automaton, a 2D depth-averaged representation of landslide runout that captures bulk transport behavior while abstracting certain physical processes, including three-dimensional flow structure, pore water pressure dynamics, rheological heterogeneity, and vegetation effects. Accordingly, all evaluations are conducted against MWR-generated trajectories, providing a controlled setting in which the underlying dynamics are known and can be systematically varied. While the study is grounded in documented landslide events, these are used to define site geometry and boundary conditions rather than to directly reproduce observed runout histories. Landslide behavior influenced by processes not represented in the model may therefore not be fully captured within the simulation ensemble, which in turn may limit the fidelity of the learned prior in such settings.

The experimental design further assumes known source polygons, duration, and failure depths; in practice, these must be estimated independently, introducing uncertainty not assessed here. 
Direct evaluation of the complete reconstructed trajectory would require temporally resolved field observations throughout the runout process, which are rarely available for rapid mass-movement events. Consequently, full-trajectory evaluation in the present study relies on
simulation-generated ground truth with withheld parameters. This provides a controlled test of the reconstruction methodology but does not fully capture the domain shift---including measurement noise, unmodeled physics, and structural discrepancies between the simulation model and real landslide mechanics---that would arise in field applications.
The validation in \cref{sec:observed_validation} provides a complementary assessment against an observed terminal deposit, with evaluation necessarily limited to the terminal state.

Moreover, the current study follows the site-specific modeling paradigm common in physics-based runout analysis, where each event is calibrated and evaluated independently. Cross-site or cross-event transfer---leveraging a model trained on one event's ensemble to reconstruct a geometrically distinct failure---is a natural next
step for establishing the operational applicability of the approach.

Finally, LAPIS-SHRED produces deterministic point estimates without explicit uncertainty quantification. Incorporating calibrated uncertainty estimates would be an important direction for extending the framework, particularly for operational decision support.

\section{Summary and Future Directions}
\label{sec:future}

We have applied LAPIS-SHRED to landslide runout reconstruction near Mount St.\ Helens, demonstrating that the framework recovers complete spatiotemporal dynamics from sparse observations under both backward (terminal-only) and forward (short initial window) inference modes.
Trained on four representative scenarios spanning dry to wet runout regimes, LAPIS-SHRED closely approaches the SHRED upper bound with full temporal observations. The method is robust to sensor placement constraints, DEM resolution choices, and the precision of prior parameter estimates, and it substantially outperforms both simple interpolation baselines and unconstrained SINDy-SHRED under the terminal-only observation constraint.

Several directions merit future investigation. First, extending the validation 
beyond the 2022 event to include field observations from additional documented landslides and simulations generated using alternative runout models is essential for establishing operational credibility. 
Second, cross-site validation would test whether a model trained on a general range of flow types at one site could transfer to sites with different terrain morphologies and failure mechanisms. Third, incorporating uncertainty quantification would provide calibrated confidence intervals for operational decision support. Fourth, coupling LAPIS-SHRED with active sensor placement optimization could inform the design of monitoring networks that maximize reconstruction accuracy under deployment constraints. Finally, integrating the framework with real-time data streams could enable continuous updating of inferences as new observations become available in practice.

\section*{Code and Data Availability}
Our codebase is available at \url{https://github.com/xswzaqnjimko/landslide_LAPIS}.

\section*{Declaration of Competing Interest}

The authors declare that they have no known competing financial interests or personal relationships that could have appeared to influence the work reported in this paper.

\section*{Acknowledgement}
YB and JNK acknowledge support from the Air Force Office of Scientific Research (FA9550-24-1-0141). EI and YB also acknowledge
support from the US National Science Foundation (NSF) CAIG program 2530591.

\appendix

\section{Scenario Parameter Specifications}
\label{app:scenarios}

\Cref{tab:scenario_params} reports the parameter ranges and ground truth configurations for all 16 scenarios derived from the 2009 Mount St.\ Helens event. Scenarios are organized by climate class (dry, moderate, wet) crossed with moisture sub-class (dryer, wetter) and soil availability level (i.e., thickness) (Shallow\,Soil: low, Deep\,Soil: high), yielding 12 climate-based scenarios. Four additional ``general site model'' scenarios share a common broad parameter range with ground truth configurations at different corners of the parameter space. The four representative scenarios evaluated in the main text are indicated with $\ast$.

For each scenario, the training ensemble of $K = 15$ simulations is generated by sampling each parameter independently and uniformly within the specified ranges. The ground truth parameter vector $\boldsymbol{\theta}_{\text{true}}$ is placed at the boundary of the ranges, lying within the convex hull of the ensemble but not identical to any training configuration.

\begin{table*}[h]
\centering
\caption{Parameter specifications for all 16 scenarios (2009 event). Ranges: uniform sampling bounds for ensemble generation. $\boldsymbol{\theta}_{\text{true}}$: ground truth test configuration. $S_c$: critical slope; $q_c$: threshold flux (m); $k$: erosion coefficient; $d_s$: soil depth (m); $d_p$: depth parameter (m). $\ast$: scenarios evaluated in main text.}
\label{tab:scenario_params}
\scriptsize
\setlength{\tabcolsep}{2.5pt}
\resizebox{\tblwidth}{!}{%
\begin{tabular}{ll ccccc ccccc}
\toprule
& & \multicolumn{5}{c}{\textbf{Parameter Range} $[\min, \max]$} & \multicolumn{5}{c}{$\boldsymbol{\theta}_{\textbf{true}}$} \\
\cmidrule(lr){3-7} \cmidrule(lr){8-12}
\textbf{Scenario} & & $S_c$ & $q_c$ & $k$ & $d_s$ & $d_p$ & $S_c$ & $q_c$ & $k$ & $d_s$ & $d_p$ \\
\midrule
\multicolumn{12}{l}{\textit{Dry climate}} \\
$\ast$\,Dryer--Shallow\,Soil & & .25--.30 & .48--.80 & .010--.015 & 1.25--1.63 & 1.25--1.75 & .30 & .80 & .010 & 1.25 & 1.25 \\
Dryer--Deep\,Soil & & .25--.30 & .48--.80 & .010--.015 & 1.63--2.00 & 1.75--3.00 & .30 & .80 & .010 & 2.00 & 3.00 \\
Wetter--Shallow\,Soil & & .20--.25 & .15--.48 & .015--.020 & 1.25--1.63 & 1.25--1.75 & .20 & .15 & .020 & 1.25 & 1.25 \\
Wetter--Deep\,Soil & & .20--.25 & .15--.48 & .015--.020 & 1.63--2.00 & 1.75--3.00 & .20 & .15 & .020 & 2.00 & 3.00 \\
\midrule
\multicolumn{12}{l}{\textit{Moderate climate}} \\
Dryer--Shallow\,Soil & & .15--.19 & .10--.14 & .021--.061 & 1.25--1.63 & 1.25--1.75 & .19 & .14 & .021 & 1.25 & 1.25 \\
Dryer--Deep\,Soil & & .15--.19 & .10--.14 & .021--.061 & 1.63--2.00 & 1.75--3.00 & .19 & .14 & .021 & 2.00 & 3.00 \\
Wetter--Shallow\,Soil & & .10--.15 & .05--.10 & .061--.100 & 1.25--1.63 & 1.25--1.75 & .10 & .05 & .100 & 1.25 & 1.25 \\
$\ast$\,Wetter--Deep\,Soil & & .10--.15 & .05--.10 & .061--.100 & 1.63--2.00 & 1.75--3.00 & .10 & .05 & .100 & 2.00 & 3.00 \\
\midrule
\multicolumn{12}{l}{\textit{Wet climate}} \\
Dryer--Shallow\,Soil & & .07--.09 & .03--.04 & .110--.180 & 1.25--1.63 & 1.25--1.75 & .09 & .04 & .110 & 1.25 & 1.25 \\
$\ast$\,Dryer--Deep\,Soil & & .07--.09 & .03--.04 & .110--.180 & 1.63--2.00 & 1.75--3.00 & .09 & .04 & .110 & 2.00 & 3.00 \\
Wetter--Shallow\,Soil & & .05--.07 & .01--.03 & .180--.250 & 1.25--1.63 & 1.25--1.75 & .05 & .02 & .250 & 1.25 & 1.25 \\
Wetter--Deep\,Soil & & .05--.07 & .01--.03 & .180--.250 & 1.63--2.00 & 1.75--3.00 & .05 & .02 & .250 & 2.00 & 3.00 \\
\midrule
\multicolumn{12}{l}{\textit{General site model} (shared range: $S_c$\,=\,.10--.15, $q_c$\,=\,.05--.80, $k$\,=\,.10--.25, $d_s$\,=\,1.25--2.00, $d_p$\,=\,1.25--3.00)} \\
High $S_c$, Low $k$ & & \multicolumn{5}{c}{(see shared range)} & .15 & .50 & .100 & 2.00 & 3.00 \\
High $S_c$, High $k$, Mid $q_c$ & & \multicolumn{5}{c}{} & .15 & .50 & .250 & 2.00 & 3.00 \\
$\ast$\,Low $S_c$, Low $k$ & & \multicolumn{5}{c}{} & .10 & .50 & .100 & 2.00 & 3.00 \\
High $S_c$, High $k$, High $q_c$ & & \multicolumn{5}{c}{} & .15 & .80 & .250 & 2.00 & 3.00 \\
\bottomrule
\end{tabular}%
}
\end{table*}

\section{Per-Scenario Reconstruction Figures}
\label{app:scenario_figures}

This appendix presents the full spatiotemporal reconstruction figures for all four representative 2009 Mount St.\ Helens scenarios evaluated in \cref{sec:results_core}. For each scenario, we show the LAPIS-SHRED reconstruction under both backward and forward inference modes, as well as the corresponding SHRED reconstruction with full temporal observations (upper bound). The Wet--Dryer--Deep\,Soil scenario is shown in the main text (\cref{fig:lapis_2009_example}); its SHRED upper bound is included here for completeness. All figures use log-scale visualization to enhance contrast for small-magnitude DoD changes; green contours outline the active region boundary.


\begin{figure}[pos=thbp]
\centering
\begin{subfigure}[t]{\columnwidth}
    \centering
    \includegraphics[width=\columnwidth,height=0.27\textheight,keepaspectratio]{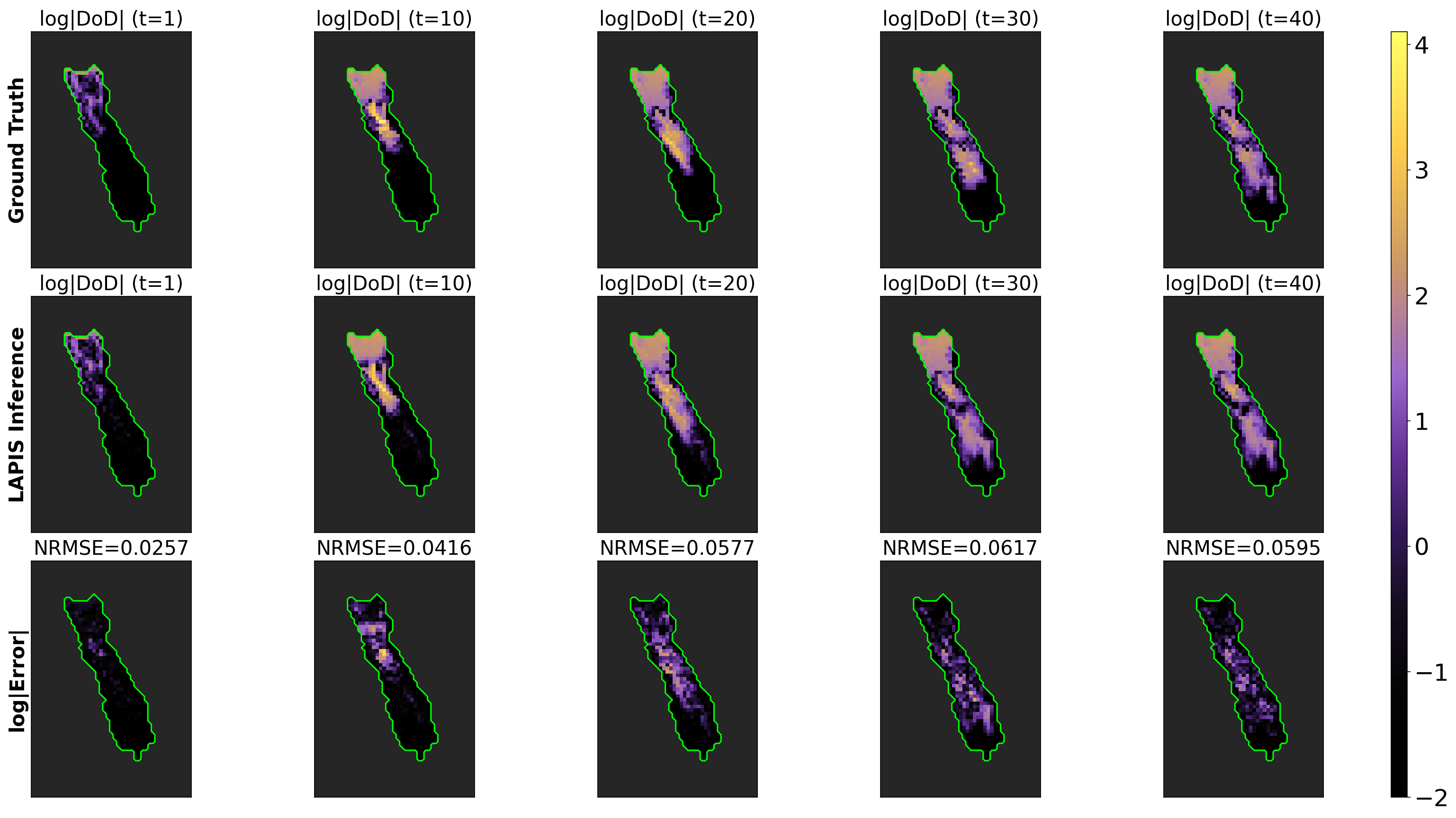}
    \caption{LAPIS-SHRED backward inference from terminal observation.}
\end{subfigure}
\\[6pt]
\begin{subfigure}[t]{\columnwidth}
    \centering
    \includegraphics[width=\columnwidth,height=0.27\textheight,keepaspectratio]{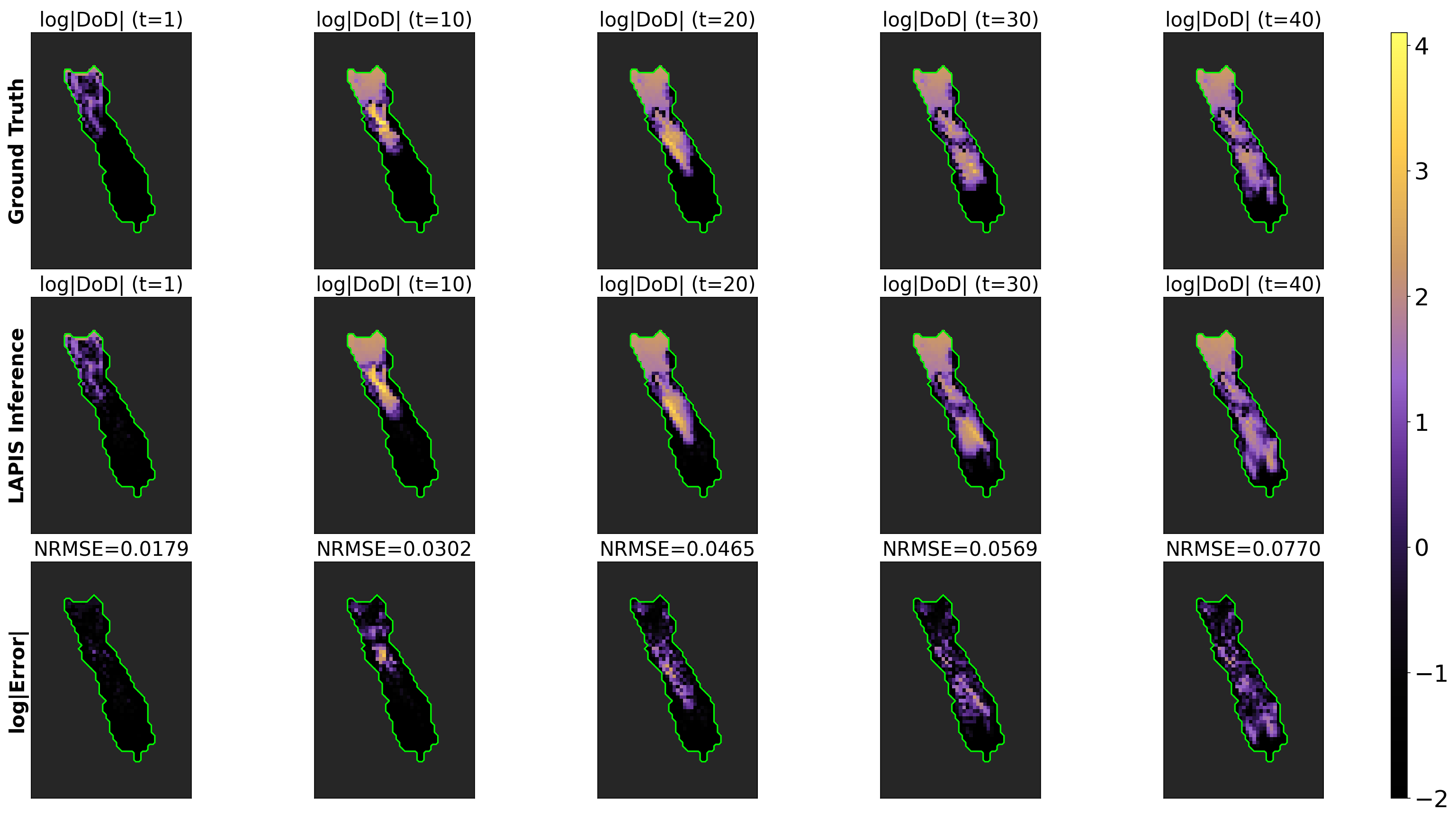}
    \caption{LAPIS-SHRED forward inference from initial observation window ($W = 5$ frames).}
\end{subfigure}
\\[6pt]
\begin{subfigure}[t]{\columnwidth}
    \centering
    \includegraphics[width=\columnwidth,height=0.27\textheight,keepaspectratio]{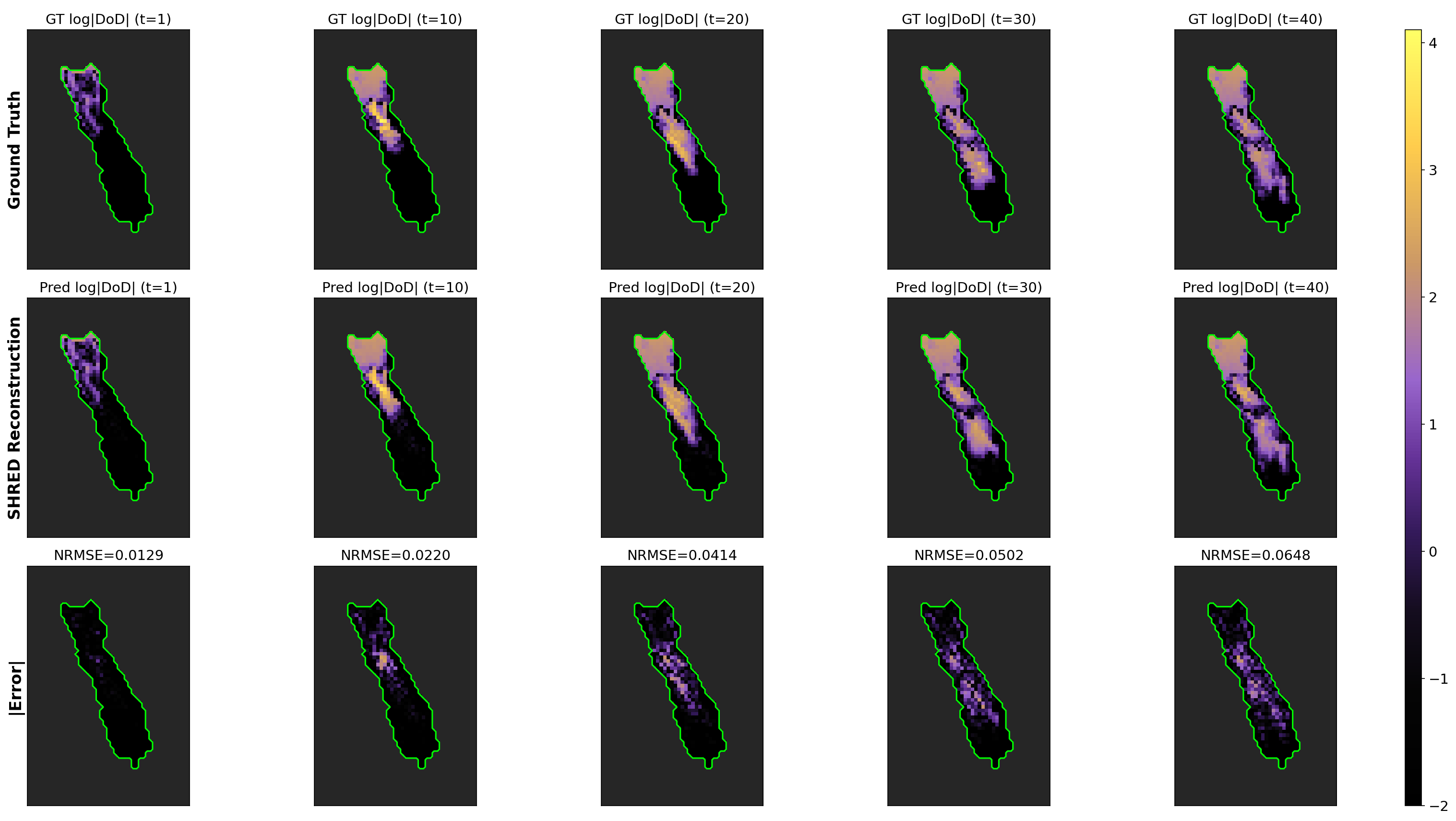}
    \caption{SHRED reconstruction with full temporal observations (upper bound).}
\end{subfigure}
\caption{Reconstruction results for the Dry--Dryer--Shallow\,Soil scenario (2009 event). Rows within each panel: ground truth DoD (top), reconstruction (middle), pointwise error with per-frame RMSE (bottom). Columns correspond to selected timesteps.}
\label{fig:app_dry_dryer_soil1}
\end{figure}


\begin{figure}[pos=thbp]
\centering
\begin{subfigure}[t]{\columnwidth}
    \centering
    \includegraphics[width=\columnwidth,height=0.27\textheight,keepaspectratio]{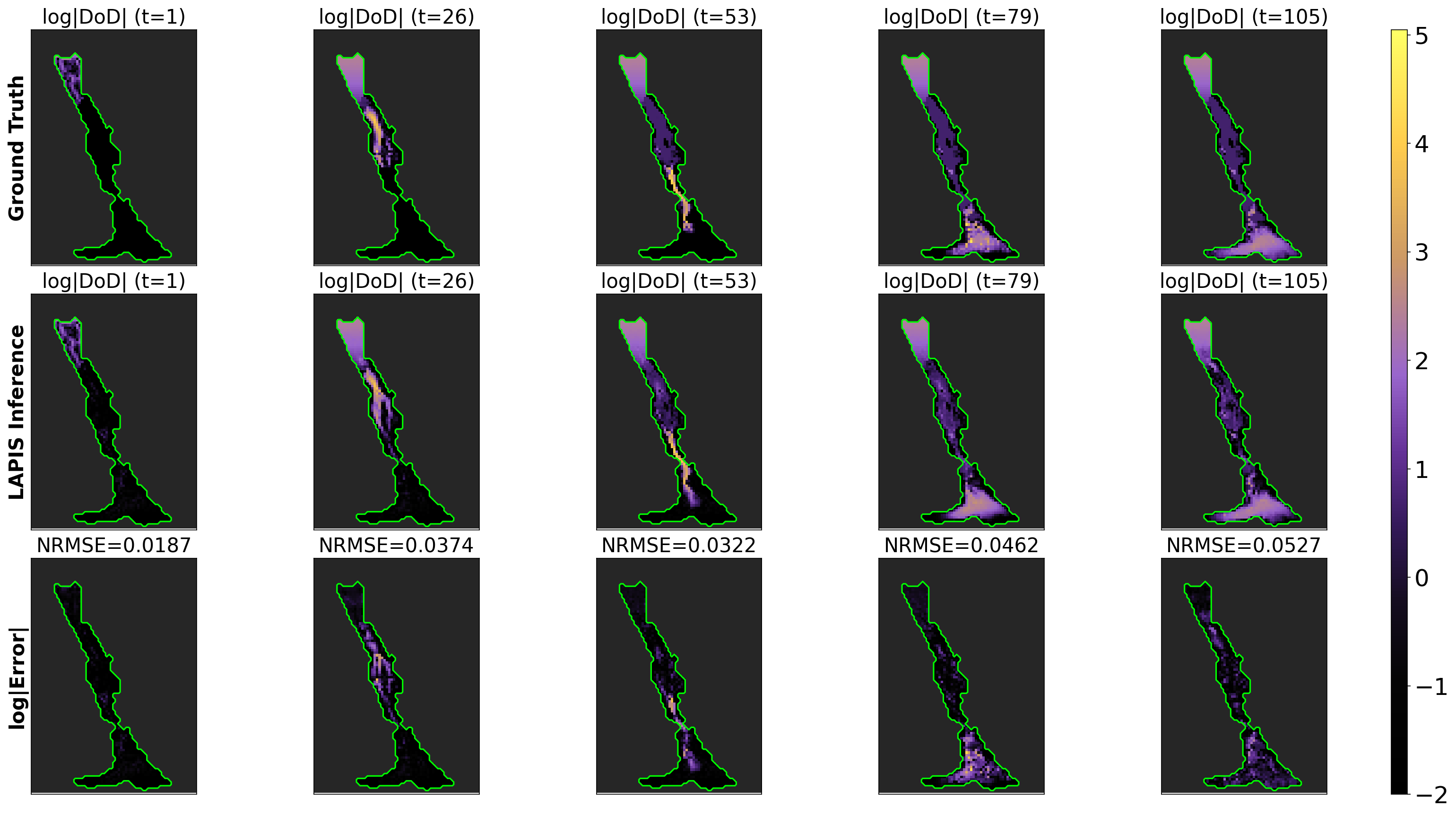}
    \caption{LAPIS-SHRED backward inference from terminal observation.}
\end{subfigure}
\\[6pt]
\begin{subfigure}[t]{\columnwidth}
    \centering
    \includegraphics[width=\columnwidth,height=0.27\textheight,keepaspectratio]{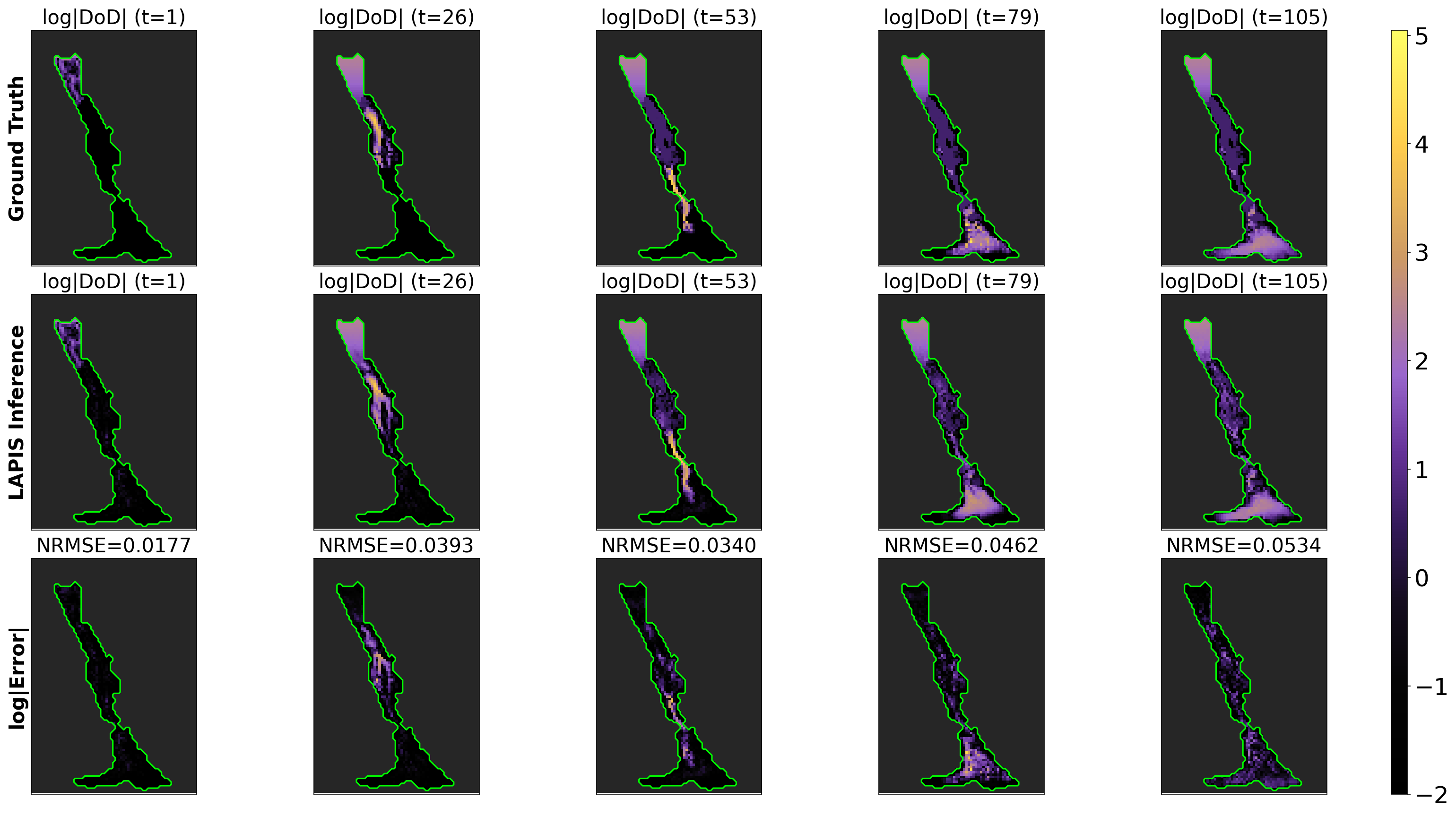}
    \caption{LAPIS-SHRED forward inference from initial observation window ($W = 10$ frames).}
\end{subfigure}
\\[6pt]
\begin{subfigure}[t]{\columnwidth}
    \centering
    \includegraphics[width=\columnwidth,height=0.27\textheight,keepaspectratio]{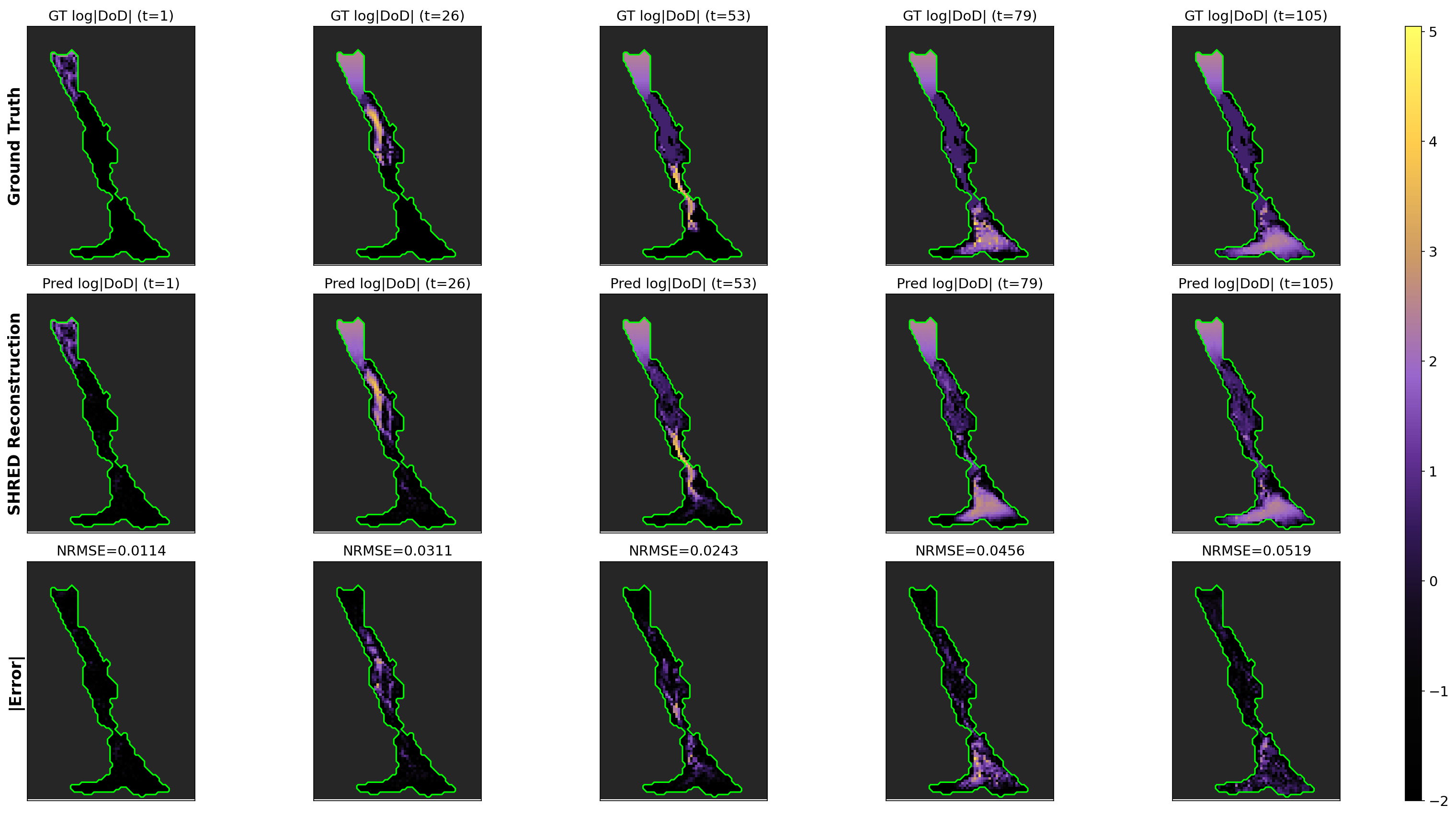}
    \caption{SHRED reconstruction with full temporal observations (upper bound).}
\end{subfigure}
\caption{Reconstruction results for the Moderate--Wetter--Deep\,Soil scenario (2009 event). Layout same as \cref{fig:app_dry_dryer_soil1}.}
\label{fig:app_moderate_wetter_soil2}
\end{figure}


\begin{figure}[pos=thbp]
\centering
\includegraphics[width=\columnwidth]{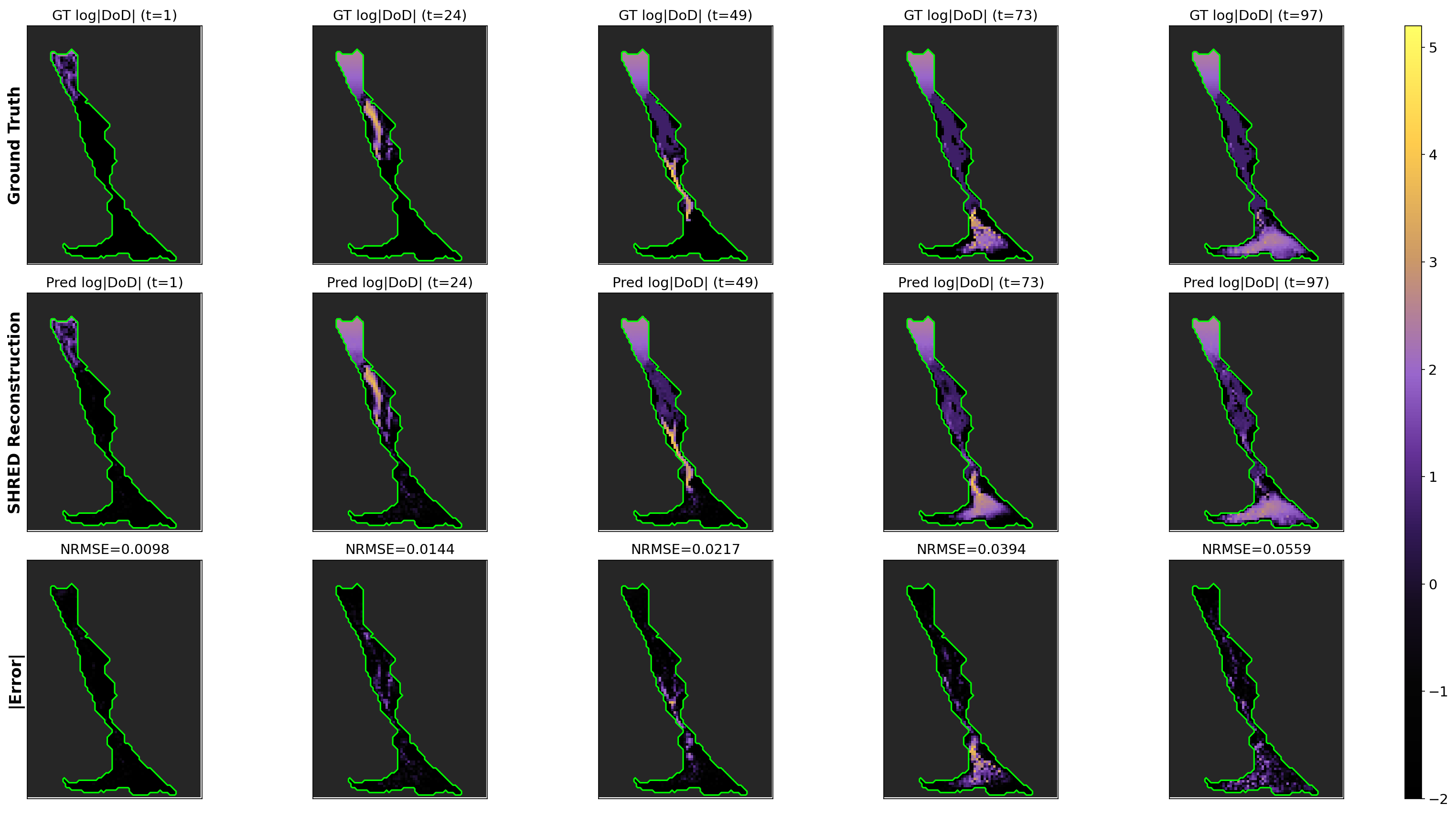}
\caption{SHRED reconstruction with full temporal observations (upper bound) for the Wet--Dryer--Deep\,Soil scenario (2009 event). The corresponding LAPIS-SHRED reconstructions under both inference modes are presented in \cref{fig:lapis_2009_example} of the main text. Layout same as individual panels in \cref{fig:app_dry_dryer_soil1}.}
\label{fig:app_wet_dryer_soil2_shred}
\end{figure}


\begin{figure}[pos=thbp]
\centering
\begin{subfigure}[t]{\columnwidth}
    \centering
    \includegraphics[width=\columnwidth,height=0.27\textheight,keepaspectratio]{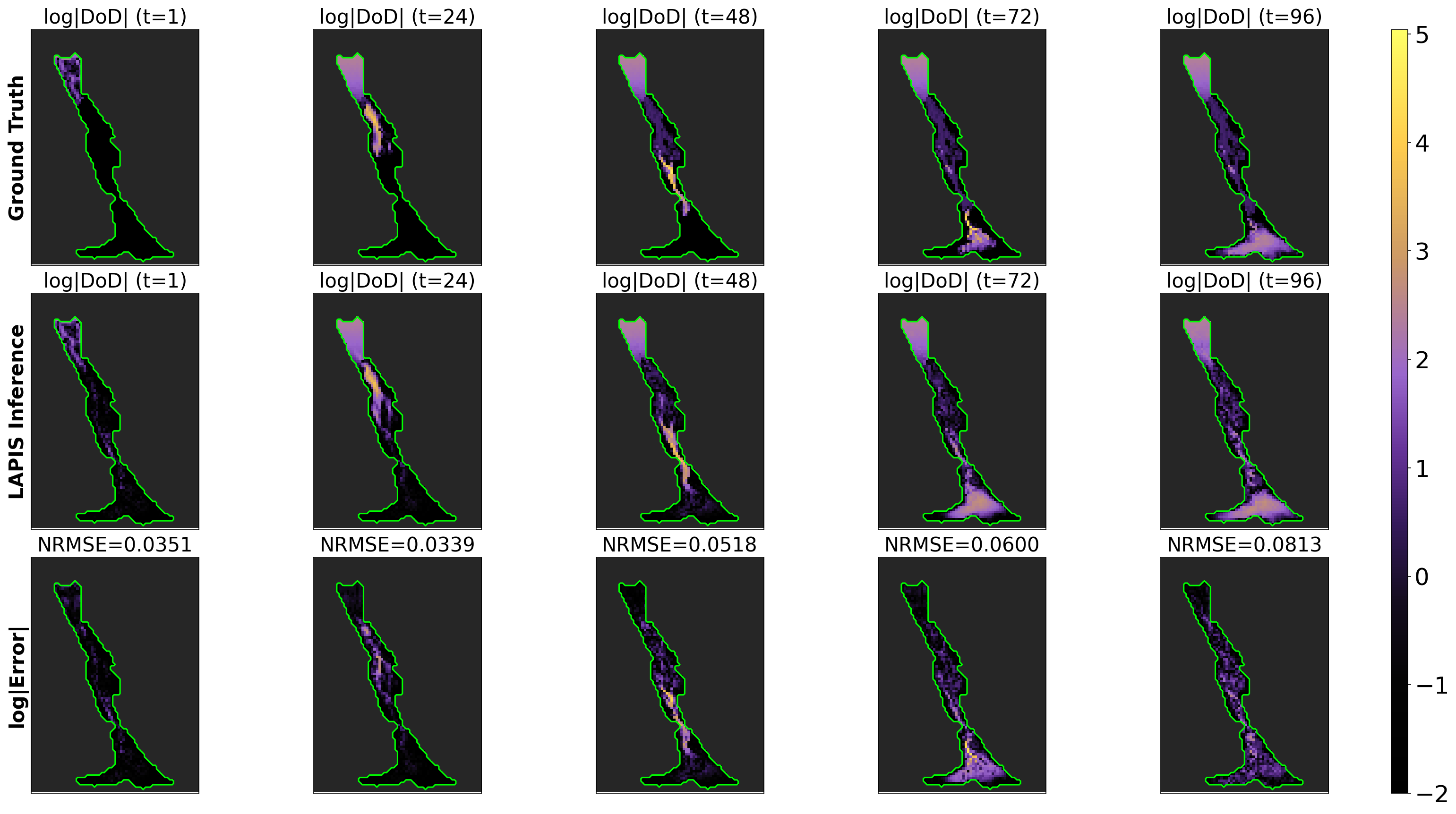}
    \caption{LAPIS-SHRED backward inference from terminal observation.}
\end{subfigure}
\\[6pt]
\begin{subfigure}[t]{\columnwidth}
    \centering
    \includegraphics[width=\columnwidth,height=0.27\textheight,keepaspectratio]{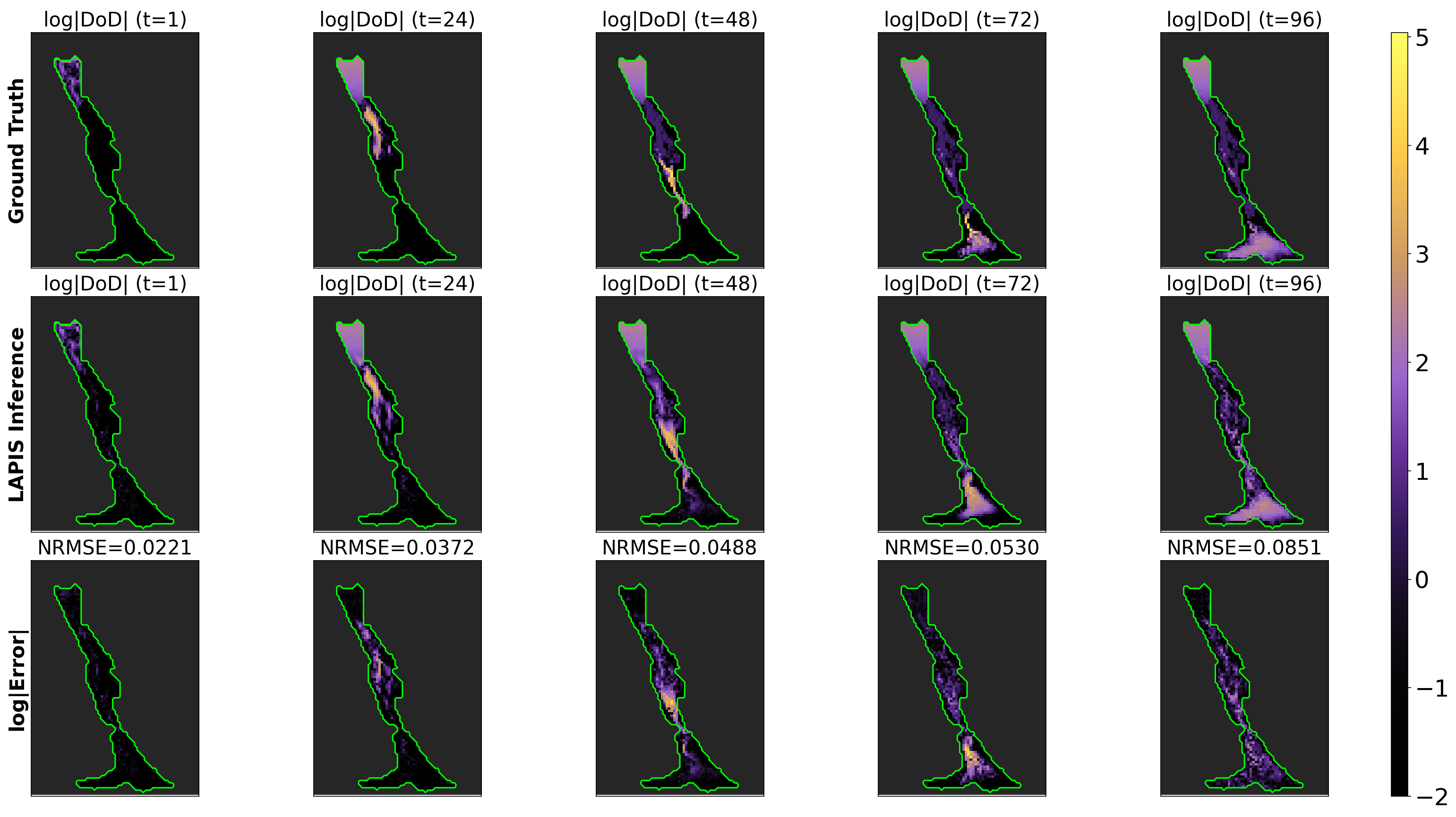}
    \caption{LAPIS-SHRED forward inference from initial observation window ($W = 9$ frames).}
\end{subfigure}
\\[6pt]
\begin{subfigure}[t]{\columnwidth}
    \centering
    \includegraphics[width=\columnwidth,height=0.27\textheight,keepaspectratio]{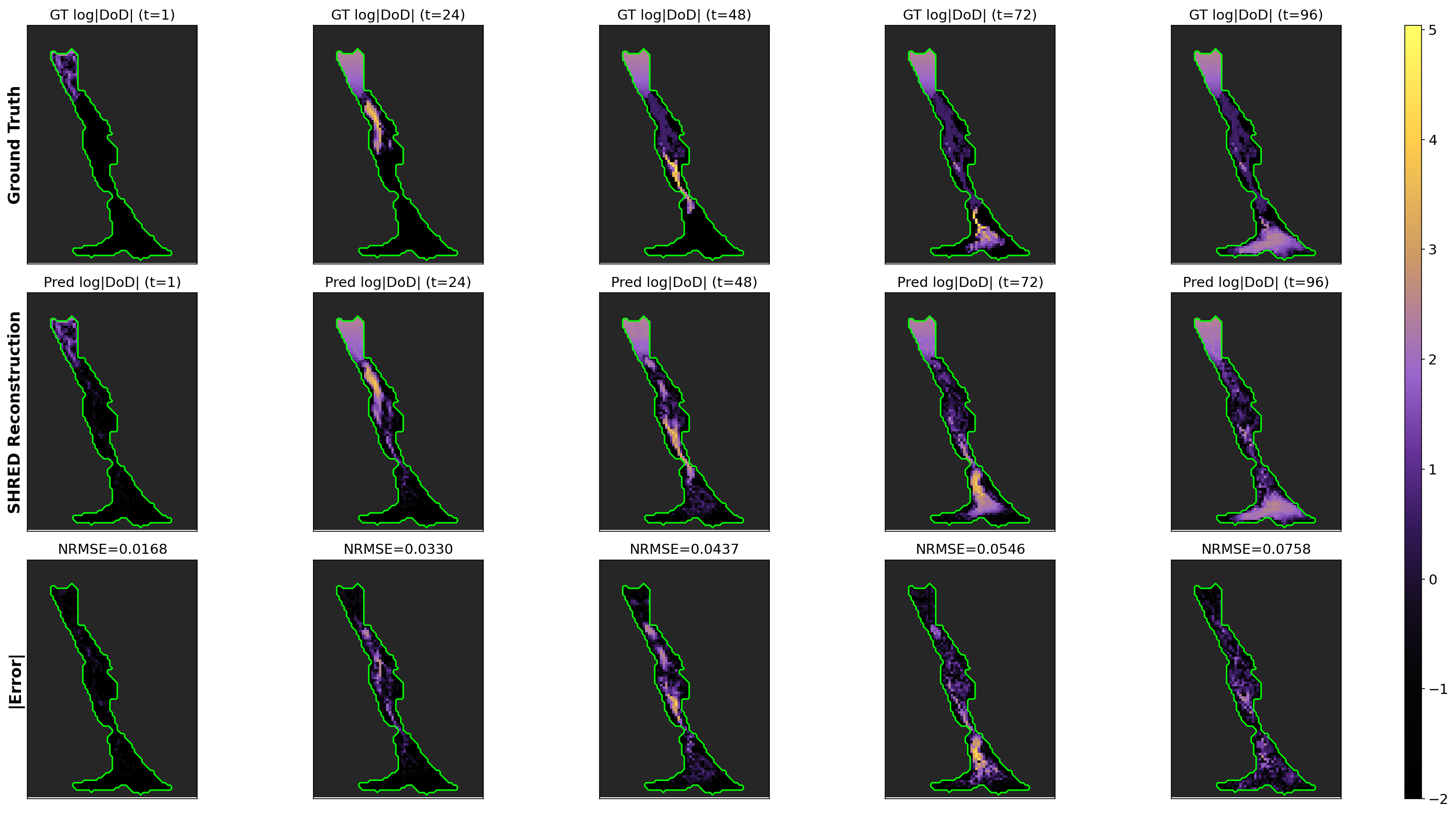}
    \caption{SHRED reconstruction with full temporal observations (upper bound).}
\end{subfigure}
\caption{Reconstruction results for the General--Low\,$S_c$--Low\,$k$ scenario (2009 event). Layout same as \cref{fig:app_dry_dryer_soil1}.}
\label{fig:app_general_low_Sc_low_k}
\end{figure}

\section{Experimental Details and Training Configuration}
\label{app:experimental_details}

This section provides supplementary details on the preprocessing, implementation, training details, data separation procedure, and the role of the shape loss weighting in the backward model training objective. These details complement the methodological exposition in the main text.

\subsection{Preprocessing Details}
\label{app:preprocessing}

Two preprocessing steps are applied to the MWR simulation outputs before they are used for training.

\textbf{Stagnation cropping.}
The MWR model iterates until all mobilized material has deposited or exited the domain, but in practice the final iterations often involve negligible boundary evolution---material has effectively ceased flowing while the solver continues executing. To restrict the analysis to the dynamically active portion of the simulation, we apply a stagnation cropping procedure. At each timestep $t$, we compare the set of active cells $\mathcal{A}_t = \{(i,j) : |\text{DoD}_t(i,j)| > \tau\}$ against the previous timestep $\mathcal{A}_{t-1}$, where $\tau = 0.1$~m is the active-cell threshold. If the number of newly activated cells $|\mathcal{A}_t \setminus \mathcal{A}_{t-1}| \leq 2$ for $W_{\mathrm{stag}}$ consecutive frames, the trajectory is truncated at the frame where stagnation began. We use $W_{\mathrm{stag}} = 5$ throughout. This cropping is applied identically to both training and ground truth simulations.

\textbf{Simulation length filtering.}
Across the parameter ensemble, extreme parameter combinations can produce simulations whose durations differ substantially from the ground truth event. Training the temporal model on such disparate sequence lengths introduces unnecessary heterogeneity in the latent trajectories. We therefore exclude training simulations whose post-crop length falls outside the interval $[T_{\mathrm{gt}} / \alpha,\; \alpha \, T_{\mathrm{gt}}]$, where $T_{\mathrm{gt}}$ is the ground truth sequence length and $\alpha = 2.5$. A minimum of three training simulations is enforced; if the filter would retain fewer, it is not applied. This step is applied only to the training ensemble; the ground truth simulation is never excluded.

\subsection{Data Separation and Evaluation Protocol}
\label{app:data_separation}

To ensure a rigorous assessment of LAPIS under the terminal-only observation constraint, we enforce strict separation between the training, inference, and evaluation phases.

During the \emph{training phase}, both SHRED and the backward model are trained exclusively on simulated trajectories $\{\mathbf{Y}^{(k)}_0, \ldots, \mathbf{Y}^{(k)}_{T_k}\}_{k=1}^K$ generated from the training parameter ensemble $\Theta_{\text{train}}$. No ground truth test data is observed at any point during training.

During the \emph{inference phase}, given a test system governed by parameters $\boldsymbol{\theta}^{\text{test}} \notin \Theta_{\text{train}}$, the model receives \emph{only} the terminal sensor observations $\mathbf{s}_T^{\text{obs}}$. The intermediate frames $\{\mathbf{Y}_1^{\text{obs}}, \ldots, \mathbf{Y}_{T-1}^{\text{obs}}\}$ are never provided to any model component. The sequence length $T$ is assumed known from event duration metadata (e.g., start and end times), which is typically available from seismic records or eyewitness reports in field settings.

During the \emph{evaluation phase}, reconstruction quality is assessed by comparing the predicted trajectory $\{\hat{\mathbf{Y}}_1, \ldots, \hat{\mathbf{Y}}_T\}$ against the held-out ground truth $\{\mathbf{Y}_1^{\text{obs}}, \ldots, \mathbf{Y}_T^{\text{obs}}\}$ using RMSE and SSIM. This evaluation data remains invisible to the model throughout the entire pipeline.

This protocol ensures that reported metrics reflect genuine reconstruction capability from terminal-only observations, without information leakage from intermediate temporal states.

\subsection{LAPIS-SHRED Implementation Details}
\label{app:implementation}

\textbf{Architecture Specifications.}
The SHRED component employs a 2-layer bidirectional LSTM with hidden dimension $d_h = 64$, yielding a latent dimension of $d_z = 128$. The decoder consists of a unidirectional LSTM with hidden dimension $d_h$, followed by a 2-layer MLP with hidden dimensions $[2d_h, 2d_h, n] = [128, 128, n]$, layer normalization, and GELU activations. The backward model adopts a 2-layer bidirectional LSTM unit and a unidirectional LSTM decoder, both with hidden dimension $d_h^{\mathcal{B}} = 64$, followed by a 2-layer output MLP with hidden dimension $2d_h^{\mathcal{B}} = 128$. Dropout of 0.1 is applied during training in both models.

\textbf{Sensor Configuration.}
We employ $p = 32$ sensors, of which approximately 18 are placed within the active region with probability proportional to DoD magnitude, and 14 are distributed across silent areas, providing static reference for reconstruction stability. The active region is defined exclusively from training simulations, thereby preventing any information leakage from test data.

\textbf{Training Details.}
SHRED is trained for 300 epochs with a batch size of 4, a learning rate of $10^{-3}$, and the AdamW optimizer with weight decay $10^{-5}$. The backward model
minimizes a composite objective over the predicted latent trajectory $\hat{\mathbf{Z}}$:
\begin{equation}
\mathcal{L}_{\text{backward}} = \lambda_r \, \mathcal{L}_{\text{recon}} + \lambda_s \, \mathcal{L}_{\text{shape}},
\label{eq:backward_loss}
\end{equation}
where $\mathcal{L}_{\text{recon}} = \frac{1}{T}\sum_{t=0}^{T-1}\|\hat{\mathbf{z}}_t - \mathbf{z}_t\|_2^2$ is the mean squared error between predicted and target latent states, and $\mathcal{L}_{\text{shape}}$ enforces structural consistency via temporal-difference matching and trajectory-variance alignment~\citep{bao2026latent}.
The backward model is trained for 500 epochs with the same optimizer configuration, using loss weights $\lambda_r = 2.0$ and $\lambda_s = 30.0$. The rationale for the elevated shape loss weight is discussed in \cref{app:shape_loss}.

\textbf{Static Padding and Ensemble Configuration.}
We use $L = 10$ padding frames, which we find sufficient for the model to reliably recognize the static terminal signature. Training employs $K = 15$ simulations spanning the physically plausible range.

\subsection{Shape Loss Weighting}
\label{app:shape_loss}

The shape loss weight $\lambda_s$ in the backward model training objective (Eq.~\ref{eq:backward_loss}) governs a trade-off between trajectory structure preservation and pointwise reconstruction accuracy. In our experiments, we adopt a relatively high weighting of $\lambda_s = 30.0$, which we find to be appropriate under the following conditions: (i) the training ensemble exhibits consistent temporal dynamics across parameter configurations, such that the trajectory ``shape'' in latent space constitutes a transferable feature; (ii) the primary objective is to recover the full progression of the physical process (e.g., initiation, peak flow, and deposition) rather than to achieve minimal pointwise error at individual timesteps; and (iii) the system exhibits characteristic temporal phases that are preserved across parameter variations.

Conversely, a lower value of $\lambda_s$ may be preferable when parameter variations induce qualitatively distinct temporal dynamics (e.g., some configurations produce sustained flow while others yield rapid deposition), rendering trajectory shape less transferable across the ensemble. A reduced $\lambda_s$ is also appropriate when pointwise reconstruction accuracy is paramount and the reconstruction loss $\mathcal{L}_{\text{recon}}$ should dominate, or when training data are limited or noisy, in which case enforcing shape matching may propagate systematic biases.

For landslide systems governed by similar underlying physics across the parameter space, the high-$\lambda_s$ regime yields superior generalization to unseen parameter configurations, as the shape loss encourages the model to learn parameter-invariant temporal structure rather than overfit to the specific dynamics of individual training trajectories.

\printcredits

\bibliographystyle{cas-model2-names}

\bibliography{refs}

\end{document}